\documentclass[prd,showpacs,floatfix,amsmath,amssymb,floatfix,nofootinbib]{revtex4} 
\usepackage{graphicx,color,dcolumn,booktabs,bm}
\usepackage{longtable,lscape,comment,braket}
\usepackage{txfonts}
\usepackage{overpic}
\usepackage{amssymb}
\usepackage{indentfirst}
\usepackage{feynmf}
\usepackage{slashed}
\usepackage{cases}
\usepackage{epstopdf}
\usepackage{psfrag}
\usepackage{subfigure}
\pdfoutput=1
\usepackage{color}
\usepackage{multirow}

\graphicspath{{Figures/}} %
\usepackage[colorlinks, citecolor=blue,anchorcolor=blue,menucolor=blue, linkcolor=blue,filecolor=blue,runcolor=blue,urlcolor=blue,frenchlinks=blue]{hyperref}
\newcommand{\be}{\begin{equation}}
\newcommand{\ee}{\end{equation}}
\newcommand{\bea}{\begin{eqnarray}}
\newcommand{\eea}{\end{eqnarray}}
\newcommand{\Z}{\mathbb{Z}} 

\def\ba{\begin{array}}
\def\ea{\end{array}}

\newcommand{\hc}[2][]{#2^{\dagger #1}} 
\newcommand{\abs}[1]{|#1|} 
\renewcommand{\Re}{\mathrm{Re}}

\newcommand{\fr}[2]{\mbox{$\frac{\,{#1}\,}{#2}$}}
\def\ie{{\it i.e.}}
\def\gev{~{\rm GeV}}

\def\lam{\lambda}

\begin{document}

\title{Dark matter and electroweak phase transition in the mixed scalar dark matter model}

\author{Xuewen Liu}
\affiliation{School of Physics, Nanjing University, Nanjing, 210093, China.}

\author{~Ligong Bian}
\email{lgbycl@cqu.edu.cn}
\affiliation{Department of Physics, Chongqing University, Chongqing 401331, China.\\
 Department of Physics, Chung-Ang University, Seoul 06974, Korea.}

\date{\today}

\begin{abstract}
We study the electroweak phase transition in the framework of scalar singlet-doublet mixed dark matter model, where the particle dark matter candidate is the lightest neutral Higgs that comprised of the CP-even component of inert doublet and a singlet scalar. The dark matter can be dominated by the inert doublet or singlet scalar depending on the mixing. We present several benchmark models to investigate the two situations after imposing several theoretical and experimental constraints. An additional singlet scalar and the inert doublet drive the electroweak phase transition to be 
strong first order. A strong first order electroweak phase transition and a viable dark matter candidate can be accomplished in two benchmark models simultaneously, where a proper mass splitting among the neutral and charged Higgs masses is needed. 

\end{abstract}



\maketitle





\section{\hspace*{-2mm}Introduction}
\label{section:introduction}

The dark matter (DM) relic abundance and baryon asymmetry of the universe (BAU) are two fundamental problems puzzling the cosmologists and  physicists. With accumulating of direct and indirect detection experiments,  
the weakly interacting massive particle (WIMP) is still one of the most  popular DM candidates, which requires new physics beyond the standard model of particle physics.
On the other hand,  the electroweak baryogenesis (EWBG) provides a very attractive mechanism to explain the BAU problem. 
One key ingredient of EWBG mechanism is 
the strong first order electroweak phase transition (EWPT), which prevents the baryon asymmetry from being washed out by the sphaleron process~\cite{Manton:1983nd,Klinkhamer:1984di,Morrissey:2012db} and helps to understand the electroweak symmetry breaking.
The model that can provide a strong first order EWPT might drive deviation of the triple Higgs coupling from the SM prediction, which can be detected at future high energy proton-proton collider~\cite{Morrissey:2012db,Arkani-Hamed:2015vfh}. It could be very intriguing if new physics can explain DM relic abundance and accomplish a strong first order EWPT.

In the simplest scalar extended dark matter models, the DM-Higgs quartic coupling leads to the interaction between the DM and nuclear.
Except the so-called {\it Higgs funnel regime} (with $m_{DM}\sim m_h/2$), the scalar singlet dark matter suffers severely constraints from the upgrade of direct detection experiments, such as 
PandaX-II~\cite{Tan:2016zwf}, LUX~\cite{Akerib:2016vxi}, and XENON~\cite{Aprile:2017iyp}, etc.
Another popular DM model is the Inert Doublet Model (IDM), where only one doublet receives a VEV in the two-Higgs doublet model. The IDM has been studied extensively as a framework to provide a scalar doublet DM~\cite{Deshpande:1977rw,Ma:2006km,Barbieri:2006dq,Gustafsson:2007pc,Agrawal:2008xz,Andreas:2009hj,Nezri:2009jd,Arina:2009um,Davoudiasl:2014pya,Gong:2012ri,Gustafsson:2012aj}. In this model, the low mass and intermediate mass regions are almost excluded by the collider constraints and the DM direct detection limits, the remnant space is the high mass region ( $>500$ GeV) which is quite difficult to probe in the direct detection experiments~\cite{Goudelis:2013uca}. Then it is very natural to consider if extending of such models can offer a possibility to avoid current and even future direct detection bounds.  
DM phenomenology aside, The strength of the EWPT 
can generally be enhanced by adding a scalar to the SM Higgs potential through cubic or quartic couplings with the SM-like Higgs field. The role played by the scalar dark matter during the universe cooling down has been studied extensively, including singlet as well as inert doublet types DM, e.g. see Ref.~\cite{Espinosa:2011eu,Espinosa:2011ax,Ghorbani:2017jls,Chen:2017qcz,Cline:2012hg,Ahriche:2012ei,Chowdhury:2011ga,
Borah:2012pu,Gil:2012ya,AbdusSalam:2013eya,Cline:2013bln,Blinov:2015vma}.

A scenario with the DM being the singlet-doublet mixed scalar is very appealing, which has been studied previously~\cite{Kadastik:2009dj,Kadastik:2009cu,Cheung:2013dua,Cohen:2011ec,Kakizaki:2016dza,Cai:2017wdu} \footnote{For the other 2HDM$+$S studies, one can see {\it e.g.} Ref.~\cite{vonBuddenbrock:2016rmr,Bell:2016ekl,Ivanov:2017dad} and the references therein.}. In such a model, the IDM is extended with a SM singlet scalar.
The inert doublet and the singlet are odd in a new $\Z_2$ discrete symmetry whereas the SM sector is even. After electroweak symmetry breaking, the electroweak charged doublet and the extra singlet mix into the scalar DM. The mixing comes from the renormalizable couplings to the SM-like Higgs field. The DM-Higgs boson triple and quartic couplings in such model are controlled by
the singlet-doublet scalars mixing. 
As a consequence, the DM annihilation and spin-independent DM-nucleon scattering processes can be characterized by the mixing. 
The thermal DM relic abundance can be achieved with an appropriate
magnitude of the mixing.
The parameter space of this kind of model has the advantage of less constrained by 
the spin-independent direct detection experiments.  

In this work, we study dark matter phenomenology and EWPT in the scalar singlet-doublet mixed DM model. Different from Refs.~\cite{Cheung:2013dua,Cohen:2011ec}, although the DM self-interactions have no impact on the DM annihilation and DM-nuclear scattering, we keep them in our model since these terms might affect the vacuum barrier during the Universe cooling down.
In the model, besides the contribution of the inert doublet to the vacuum barrier, there is an additional contribution comes from the singlet scalar which is also determined by the mixing. Therefore, a proper mixture also characterizes the realization of the strong first order EWPT. 
In this model, we find that the additional singlet along with the mixing effects  make the situation of EWPT different from the IDM, 
two benchmarks with moderate dark matter mass are allowed to gain the correct relic abundance and to achieve a strong first order EWPT. Some features of the model are found and listed bellow:

\begin{enumerate}

\item 
The vacuum stability and $T$ parameter constrain the singlet-doublet mixing non-trivially.

\item
DM relic density and the DM-nuclear scattering cross section highly depend on the mixing of the singlet and doublet scalars.

\item
Co-annihilation effects play an important role near the degenerate masses region of the dark sector, which lead to the cancellations between the annihilation processes to yield correct DM relic density. 

\item  
A proper mass splitting between the $\Z_2$ odd  neutral Higgs and charged Higgs is needed to obtain a strong first order EWPT.

\end{enumerate}

The paper is organized as follows. In Section~\ref{S:SIDM}, we present the mixed scalar dark matter model. 
Theoretical constraints and LEP bounds on scalar masses are addressed in Section~\ref{sec:cons}. The Section~\ref{dir-det} is devoted to the comments on 
DM relic density and direct detection in the model.
 In the Section~\ref{sec:PT}, we present the approach of estimation of the strength of EWPT with thermal effective potential. 
In Section~\ref{sec:DMP}, we give a comprehensive analysis of DM phenomenology and EWPT in several benchmark models. After that, we address the collider prospects of the benchmark models in Section~\ref{sec:cllider}. We conclude in the Section~\ref{sec:con}.


\section{The mixed singlet-doublet scalar dark matter model}
\label{S:SIDM}

We extend the Inert Doublet Model with an additional real singlet scalar ($S$) and assume the $\Z_2$ symmetry to inert doublet $\Phi_I$ and the $S$, the tree level potential of the model is 
\bea\label{eq:tre}
V &=& \mu_{H}^2 {\Phi_H}^\dagger {\Phi_H} + \mu_{I}^2  {\Phi_I}^\dagger {\Phi_I} + 
\mu_S^2 S^2 + \lambda_1 ({\Phi_H}^\dagger {\Phi_H})^2
+ \lambda_2 ({\Phi_I}^\dagger {\Phi_I})^2 + \lambda_3
(\Phi_H^\dagger \Phi_H)(\Phi_I^\dagger \Phi_I) \nonumber \\ &+&
\lambda_4 (\Phi_I^\dagger \Phi_H)(\Phi_H^\dagger \Phi_I) + {1 \over 2} \lambda_5
[(\Phi_I^\dagger \Phi_H)^2 + (\Phi_H^\dagger \Phi_I)^2] + \lambda_6 S^4 +  \lambda_7 S^2 (\Phi_H^\dagger \Phi_H) \nonumber\\&+&  \lambda_8
S^2 (\Phi_I^\dagger \Phi_I)+\mu_{\text{soft}}  S (\Phi_H^\dagger \Phi_I + \Phi_I^\dagger \Phi_H)\, ,
\eea  
where ${\Phi_H}$ is the SM Higgs doublet and ${\Phi_I}$ is the 
inert doublet. 
The doublets are given as
\bea
\Phi_H = \left( \begin{array}{c}
                           G^+  \\
        \frac{1}{\sqrt{2}}(v+h+i G^0)  
                 \end{array}  \right) \, ,                     
&& \Phi_I =\left( \begin{array}{c}
                           H^+   \\
        \frac{1}{\sqrt{2}}(H_0+iA)  
                 \end{array}  \right) \, ,
\label{eq:fd}
\eea 
where $\Phi_H$ develops a VEV $v=246$ GeV whereas the inert doublet and the real singlet scalar $S$ do not produce 
any VEV. The $\mathbb{Z}_2$ symmetry of $\Phi_I, S$ remains unbroken. Goldstones $G^+$ and $G^0$ are absorbed in $W^\pm$, $Z$ bosons after 
spontaneous symmetry breaking. 
All the parameters in
Eq.~\ref{eq:tre} are assumed to be real.   
By minimizing the potential in Eq.~\ref{eq:tre} we obtain the relation
\bea
\mu_{H}^2=-\lambda_1 v^2 \, ,
\label{15}
\eea 
and the mass terms for the SM Higgs and the charged scalars can be obtained as
\begin{eqnarray}
m_h^2&=&2 \lambda_1 v^2 \; ,\nonumber \\
m_{H^{\pm}}^{2}&=&\mu_{I}^{2}+\lambda_{3}\frac{v^{2}}{2}\; .
\label{16}   
\end{eqnarray}

Working in  the basis of real neutral scalars, $( S, H_0, A )$, where  $H_0$ and $A$ are the real and imaginary components of the
neutral component of $\Phi_I$. The mass squared matrix is
\begin{equation}
\mathbf{M}^2 = \left(\begin{array}{ccc} 2 \mu_S^2 +  v^2
  \lambda_7 &  v\mu_{\text{soft}} & 0 \\
  v\mu_{\text{soft}}&  \mu_I^2 +  v^2 \left( \lambda_3 + \lambda_4 +
\lambda_5 \right)/2 & 0 \\
0 & 0 &  \mu_I^2  + v^2 \left(\lambda_3 + \lambda_4 -
\lambda_5  \right)/2
\end{array} \right).
\label{mix}
\end{equation}
Considering one of the mixed states of the CP-even scalars $S$ and $H_0$ is the lightest particle among  the  three eigenstates, then it could serve as the DM candidate, and the DM mixing is induced by the upper left $2\times 2$ block of the mass matrix Eq.~\ref{mix}. The custodial SU(2) symmetry could be violated slightly by the mixing of $S$ and $H_0$, which is important for the $T$ parameter exclusion and therefore the choice of the parameters when we explore the EWPT and dark matter phenomenology. Usually the relation between $S,H_0$ and mass eigenstates $\chi$ and $H$ can be written as
\begin{align}
&\quad\left(\begin{array}{c}
    S \\
    H_0 \\
  \end{array}\right)
=\left(  \begin{array}{cc}
    \cos\theta & -\sin\theta \\
    \sin\theta & ~\cos\theta \\
  \end{array}\right)
\left(\begin{array}{c}
    \chi \\
    H \\
  \end{array}\right),
\end{align}
with the mixing angle
\begin{align}
\theta=\frac{1}{2}\tan^{-1}(\frac{2v \mu_{\text{soft}}}{\tilde{M}_S^2-\tilde{M}_{H_0}^2}),
\end{align}
and the two eigenvalues of the mass-squared
matrix are
\begin{eqnarray}
\frac{1}{2} \left[ \tilde{M}_S^2 + \tilde{M}_{H_0}^2 \pm \sqrt{\left(
    \tilde{M}_S^2 - \tilde{M}_{H_0}^2 \right)^2 + 4 v^2 \mu_{\text{soft}}^2} \right]\,,
\end{eqnarray}
where $\tilde{M}_S^2 =2 \mu_S^2 + v^2 \lambda_7 $ and
$\tilde{M}_{H_0}^2 = \mu_I^2 +  v^2 \left( \lambda_3 + \lambda_4
+ \lambda_5 \right)/2 $.  Either $\chi$ or $H$ can be the DM candidate. 
 
The masses and interactions of the scalar sector are parameterized by the scalar-potential parameters $\lambda_{1...8}$ and $\mu_{H,I,S,\text{soft}}$, which
can be traded for five physical masses
$\{m_h, m_\chi, m_H, m_{A}, m_{H^{\pm}}\}$ along with $\{\lambda_L,\lambda_{2,6,7,8},\theta \}$,
where $\lambda_L=(\lambda_3+\lambda_4+\lambda_5)/2$. 
Then, the other parameters in the scalar potential in terms of the physical ones are expressed as
\begin{eqnarray}
\lambda_3 & = & \frac{2}{v^2}(-m_H^2 \cos^2\theta + m_{H^{\pm}}^2 -m_\chi^2 \sin^2\theta + \lambda_L v^2)\; , \\
\lambda_4 & = & \frac{1}{v^2}(m_{A}^2 + m_H^2 \cos^2\theta - 2 m_{H^{\pm}}^2 + m_\chi^2 \sin^2\theta)\; , \\
\lambda_5 & = & \frac{1}{v^2}(-m_{A}^2 + m_H^2 \cos^2\theta + m_\chi^2 \sin^2\theta)\; , \\
\mu_S^2   & = & \frac{1}{2}(m_\chi^2 \cos^2\theta +m_H^2 \sin^2\theta  -\lambda_7 v^2)\; , \\
\mu_I^2   & = &  m_H^2 \cos^2\theta + m_\chi^2 \sin^2\theta -\lambda_L v^2\; , \\
\mu_{\text{soft}} & = &\frac{1}{2v}\left(m_{\chi}^2-m_{H}^2\right)\sin2\theta\; .
\label{eq:scalarcouplings}
\end{eqnarray}
The associated DM-Higgs couplings that characterize the the Higgs portal annihilation of the DM and the DM-nucleon scattering are given by 
\begin{eqnarray}
a_{h\chi\chi} = 2 v \lambda_7 \cos^2\theta + 2 v \lambda_L \sin^2\theta + \mu_{\text{soft}} \sin 2 \theta  \, ,
\label{eq:hxx}
\end{eqnarray}
or
\begin{eqnarray}
a_{hHH}  = 2 v \lambda_7 \sin^2\theta + 2 v \lambda_L \cos^2\theta - \mu_{\text{soft}} \sin 2 \theta  \,.
\label{eq:hHH}
\end{eqnarray}

\section{Theoretical  and Experimental constraints}
\label{sec:cons}

\subsection{Perturbative unitarity}
\label{sec:unitarity}

In the high energy limit, the quartic contact interaction terms contribute to the tree-level scalar-scalar scattering matrix dominantly. The $s$-wave scattering amplitudes are constrained by the perturbative unitarity limits, which requires that the eigenvalues of the $S$-matrix $\mathcal{M}$ must be smaller than the unitarity bound:
\begin{equation}
  \abs{\Re \mathcal{M}} < \frac{1}{2}.
\end{equation}

The perturbative unitarity of the general two-Higgs-Doublet model were first studied in Ref.~\cite{Kanemura:1993hm,Akeroyd:2000wc}. In this work we extend the formalism of Ref.~\cite{Ginzburg:2003fe,Ginzburg:2005dt} for the IDM  to states containing an extra singlet $S$. The initial states are classified according to their total hypercharge $Y$ ($0$, $1$ or $2$), weak isospin $\sigma$ ($0$, $\frac{1}{2}$ or $1$) and discrete $\Z_2$ charge $X$. 

For simplicity, we only list here the initial states with hypercharge $Y = 0$ and $\sigma = 0$ which differ from the 2HDM initial states~\cite{Ginzburg:2003fe,Ginzburg:2005dt}:
\begin{equation}
 \frac{1}{\sqrt{2}} \hc{\Phi_{H}} \Phi_{H}, 
 \frac{1}{\sqrt{2}} \hc{\Phi_{I}} \Phi_{I}, 
 \frac{1}{\sqrt{2}} S^{2}, 
 \frac{1}{\sqrt{2}} \hc{\Phi_{H}} \Phi_{I}, 
 \frac{1}{\sqrt{2}} \hc{\Phi_{I}} \Phi_{H},
\end{equation}
where the first three states are even under $\Z_{2}$ and the last two states are odd.
In the following, we present all the scattering matrices of the model:
\begin{align}
  8 \pi S_{Y=2,\sigma=1} &= 
  \begin{pmatrix}
    2 \lambda_{1} & \lambda_{5} & 0 \\
    \lambda_{5}^{*} & 2 \lambda_{2} & 0 \\
    0 & 0 & \lambda_{3} + \lambda_{4}
  \end{pmatrix},
  &
  8 \pi S_{Y=2,\sigma=0} &= \lambda_{3} - \lambda_{4},
  \\
  8 \pi S_{Y=0,\sigma=1} &= 
  \begin{pmatrix}
    2 \lambda_{1} & \lambda_{4} & 0 & 0 \\
    \lambda_{4} & 2 \lambda_{2} & 0 & 0 \\
    0 & 0 & \lambda_{3} & \lambda_{5}^{*} \\
    0 & 0 & \lambda_{5} & \lambda_{3}
  \end{pmatrix},
  &
  8 \pi S_{Y=1,\sigma=1/2} &= 
    \begin{pmatrix}
      2 \lambda_{7} & 0  \\
      0 & 2 \lambda_{8}    
    \end{pmatrix},
\end{align}
\begin{equation}
    8 \pi S_{Y=0,\sigma=0} = 
  \begin{pmatrix}
    6 \lambda_{1} & 2 \lambda_{3} + \lambda_{4} & \sqrt{2} \lambda_{7} & 0 & 0  \\
    2 \lambda_{3} + \lambda_{4} & 6 \lambda_{2} & \sqrt{2} \lambda_{8} & 0 & 0  \\
    \sqrt{2} \lambda_{7} & \sqrt{2} \lambda_{8} & \lambda_{6} & 0 & 0  \\
      0 & 0 & 0  & \lambda_{3} + 2 \lambda_{4} & 3 \lambda_{5}^{*}   
    \\
    0 & 0 & 0 & 3 \lambda_{5} & \lambda_{3} + 2 \lambda_{4} \\
  \end{pmatrix}.
\end{equation}

Then the eigenvalues $\Lambda_{Y\sigma i}^{X}$ of the above scattering matrices (where $i = \pm \text{ or }1,2,3$) can be calculated as
\begin{align}
  \Lambda_{21\pm}^{even} &= \lambda_{1} + \lambda_{2} \pm \sqrt{(\lambda_{1} - \lambda_{2})^{2} + \abs{\lambda_{5}}^{2}},~
  \Lambda_{21}^{odd} = \lambda_{3} + \lambda_{4}\;,~
  \Lambda_{20}^{odd} = \lambda_{3} - \lambda_{4}\; ,  \\
   \Lambda_{01\pm}^{even} &= \lambda_{1} + \lambda_{2} \pm \sqrt{(\lambda_{1} - \lambda_{2})^{2} + \lambda_{4}^{2}}\;,\qquad
  \Lambda_{01\pm}^{odd} = \lambda_{3} \pm \abs{\lambda_{5}},
  \\
    \Lambda_{00\pm}^{odd} &= \lambda_{3} +2\lambda_{4}\pm 3\abs{\lambda_{5}},
\label{eq:Z2:unitarity}
\end{align}
and the $\Lambda_{00\, 1,2,3}^{even}$ correspond to  
the three roots of the polynomial equation
\begin{equation}
\begin{split}
  & x^3 - x^2 (6 \lambda_{1} + 6 \lambda_{2} +  \lambda_{6}) + x (36\lambda_{1} \lambda_{2}-4\lambda_{3}^2-4\lambda_{3} \lambda_{4}- \lambda_{4}^2+6\lambda_{1} \lambda_{6}+6\lambda_{2} \lambda_{6} \\
   &- 2\lambda_{7}^2 - 2\lambda_{8}^2) 
   +12 \lambda_{1} \lambda_{8}^2 +
 12 \lambda_{2} \lambda_{7}^2 - 4\lambda_{4}\lambda_{7}\lambda_{8}- 8\lambda_{3}\lambda_{7}\lambda_{8}+\lambda_{4}^2\lambda_{6}+4\lambda_{3}\lambda_{4}\lambda_{6}\\
 &+4\lambda_{3}^2\lambda_{6}-36\lambda_{1}\lambda_{2}\lambda_{6}
 = 0.
\end{split}
\end{equation}

\subsection{Vacuum stability conditions }

To make sure the potential is bounded from below in the limit of large field values, we impose copositivity criteria on the quartic couplings of the tree level potential as explored in Ref.~\cite{kannike}. 
The vacuum stability conditions for the model are given as
\begin{align}
  \lambda_{1} \; ,  ~\lambda_{2}\;  ,~ \lambda_{6} > 0\; , ~~~~~~~~~
2 \sqrt{\lambda_{1} \lambda_{6}} + \lambda_{7} > 0 ,~~~~~~~~~
  2 \sqrt{\lambda_{2} \lambda_{6}} + \lambda_{8} > 0\; , 
  \end{align}
  and for $\lambda_4>0$ case,
  \begin{align}
&2 \sqrt{\lambda_{1} \lambda_{2}} + \lambda_{3} > 0\; ,\\ 
   & \sqrt{\lambda_{1} \lambda_{2} \lambda_{6}} 
  + \sqrt{\lambda_{1}} \lambda_{8} + \sqrt{\lambda_{2}} \lambda_{7}
  + \sqrt{\lambda_{6}} \lambda_{3}\notag \\
  &+ \sqrt{(2 \sqrt{\lambda_{1} \lambda_{6}} + \lambda_{7}) (2 \sqrt{\lambda_{2} \lambda_{6}} + \lambda_{8})
  (2 \sqrt{\lambda_{1} \lambda_{2}} + \lambda_{3})} > 0\; ,
    \end{align}
  if $\lambda_4<0$, one obtains  
  \begin{align}
  &2 \sqrt{\lambda_{1} \lambda_{2}} + \lambda_{3} + \lambda_{4} - |\lambda_{5}| > 0\; , \\
  &   \sqrt{\lambda_{1} \lambda_{2} \lambda_{6}} 
  + \sqrt{\lambda_{1}} \lambda_{8} + \sqrt{\lambda_{2}} \lambda_{7}
  + \sqrt{\lambda_{6}} (\lambda_{3} + \lambda_{4} - |\lambda_{5}|) \notag\\
  &+ \sqrt{(2 \sqrt{\lambda_{1} \lambda_{6}} + \lambda_{7}) (2 \sqrt{\lambda_{2} \lambda_{6}} + \lambda_{8})
  (2 \sqrt{\lambda_{1} \lambda_{2}} + \lambda_{3} + \lambda_{4} - |\lambda_{5}|)} > 0\; .
\end{align}

It should be noted that, with loop corrections being included, one may expect more viable parameter spaces allowed by the vacuum stability conditions.

\subsection{Electroweak precision test}

We should also make sure that 
electroweak precision test data is respected when we study the phenomenology of the model.
As we know, electroweak precision test can be expressed in terms of three measurable quantities, called $S, T$ and $U$, which parameterize the contributions
from beyond standard model physics  to electroweak radiative corrections~\cite{Peskin:1991sw}.
The most relevant one is the $T$ parameter, which characterize the spoil extent of the global $SU(2)$ symmetry. 
Since the scalar singlet $S$ mixes with the $H_0$ in this model, the self-energies of the Goldstone bosons get corrected by the diagrams with virtual inert particle pairs and $S$. Following the method in Ref.~\cite{Barbieri:2006dq}, we first obtain
\begin{align}\label{rho1}
\Delta\rho &  =\frac{(\lambda_{4}+\lambda_{5})^{2}}{2}\left(\cos^2\theta f(m_{_{H^\pm}},m_{H})+\sin^2\theta f(m_{_{H^\pm}},m_{\chi})\right)\nonumber\\
&+\frac{(\lambda
_{4}-\lambda_{5})^{2}}{2}f(m_{_{H^{\pm}}},m_{A})-2(\lambda_{5}\cos\theta-\mu_{\text{soft}}\sin\theta/v)^2 f(m_{A},m_{H})\nonumber\\
&-2(\lambda_{5}\sin\theta+\mu_{\text{soft}}\cos\theta/v)^2f(m_{A},m_{\chi})\; ,\\
\nonumber\\
\mathrm{with}&~~~ f(m_{1},m_{2})    =\frac{v^{2}}{32\pi^{2}}\int_{0}^{1}\frac{dx\,x(1-x)}%
{xm_{1}^{2}+(1-x)m_{2}^{2}}\; ,
\end{align}

In the limit of custodial SU(2) symmetry, i.e., $m_{H^\pm}=m_A$,  
Eq.~\ref{rho1} can be recast as 
\begin{align}\label{rho2}
\Delta\rho &  =\frac{(\lambda_{4}+\lambda_{5})^{2}}{2}\left(\cos^2\theta f(m_{_{H^\pm}},m_{H})+\sin^2\theta f(m_{_{H^\pm}},m_{\chi})\right)\nonumber\\
&-2(\lambda_{5}\cos\theta-\mu_{\text{soft}}\sin\theta/v)^2 f(m_{A},m_{H})\nonumber\\
&-2(\lambda_{5}\sin\theta+\mu_{\text{soft}}\cos\theta/v)^2f(m_{A},m_{\chi})\; ,
\end{align} 
which sets bounds on the magnitude of the mixing between $S$ and $H_0 $, as well as the mass splitting among neutral and charged Higgses in the model.
The $T$ parameter can be obtained as $T=\Delta\rho/\alpha_{EM}$ following the notation of Ref.~\cite{Barbieri:2006dq}.
The Gfitter fit to the electroweak data~\cite{Baak:2014ora}: $T=0.09\pm 0.13$ has been used to constrain the parameters.

\subsection{LEP bounds}

The LEP bounds on the scalar masses of this model can be considered as the same as in IDM,  and the only difference is that the CP-even neutral Higgs $H$ is an admixture of doublet and singlet.  
The precise measurements of the $W$ and $Z$ widths lead to the following lower limits on the  scalar masses
\begin{eqnarray}
\label{eq:constr-widths}
&& m_{H/\chi} + m_{H^{\pm}} > m_{W^{\pm}}\;, \quad \quad m_{A} + m_{H^{\pm}} > m_{W^{\pm}} \; ,\nonumber\\
&& m_{H/\chi} + m_{A} > m_{Z}\; , \quad  \quad 2m_{H^{\pm}} > m_{Z}\; ,
\end{eqnarray}
to ensure that decay channels of ${W^{\pm} \rightarrow H/\chi H^{\pm},AH^{\pm}}$ and ${Z \rightarrow H/\chi A,H^+H^-}$ are kinematically 
forbidden. The production of the charge Higgs pairs $e^+e^-\to H^+H^-$ at LEP-II~\cite{Pierce:2007ut}
sets
\begin{equation}
m_{H^\pm}>70\mbox{ GeV}\; ,
\label{eq:mhcp-lep2}
\end{equation}
which does not apply when the scalar mass is larger than $m_Z/2$ providing the mass splitting $m_{H^\pm}-m_H$ is smaller than 5 GeV~\cite{Blinov:2015qva}.
For the neutral Higgses of the model, the constraints come from the pair production process of $e^+e^-\rightarrow H/\chi A $ followed by the cascade decay $A\rightarrow H/\chi Z\rightarrow H/\chi f\bar{f}$, which can be obtained through a reinterpretation of the neutralino production search at LEP-II. The analysis of Ref.~\cite{Lundstrom:2008ai} shows that the limit on the neutral Higgs is $max$($m_A$,$m_H$)$\geq 100$ GeV for the IDM, which can be roughly applied to our model when the mixing of $H_0$ and $S$ is small, or can be interpreted as $max$($m_A$,$m_\chi$)$\geq 100$ GeV when the mixing of $H_0$ and $S$ is large, which can be figured out from Eq.~\ref{eq:lc} in the Section ~\ref{sec:LPP} .

\section{DM Relic density and Direct Detection}
\label{dir-det}
The thermal DM relic density should match the observed DM density in the Universe of $\Omega h^2 = 0.1199\pm0.0022$~\cite{Ade:2015xua}, or at least of not over-producing such density via thermal production. To an approximation, the relic density is given by
\begin{equation}
\Omega_{DM} 
\simeq 1.07\times10^9  \frac{x_f}{\sqrt{g_*} \, M_{\rm{Pl}}\langle \sigma_{\rm ann} v_{\rm rel} \rangle} \gev^{-1}\; ,
\end{equation}
where $x_f = m_{D}/T_f \simeq 20$ with $T_f$ being  the typical freeze-out temperature of a WIMP~\cite{Kolb:1990vq}, $m_D=m_\chi$ or $m_H$ is the DM mass, $M_{Pl}$ is the Planck mass, $g_*$ is number of relativistic degrees of freedom,  $\langle \sigma v \rangle$ is the thermally averaged cross section for DM pair annihilation into the SM particles (\ie\  $f\bar{f}$, $W^+W^-, ZZ, hh$). 
For the mixed DM scenario, the DM annihilation is characterized by the masses of dark matter and the mixing of DM constituents.
To illustrate the DM annihilation properties with different channels in this model, we collect several benchmarks in distinct DM mass regions in the following subsection. 
The relic density is calculated by implementing package {\tt  micrOMEGAs}~ \cite{Belanger:2014vza}.

In the direct detection experiment of dark matter, the nuclear recoil spectrum is directly related to the DM-nuclei scattering cross-section~\cite{Jungman:1995df}, which is given by:
\begin{equation}
\sigma_{ D-nucl} = \int_{0}^{4 \mu_r^2 v^2} \frac{d \sigma (q=0)}{d|\textbf{q}|^2} d|\textbf{q}|^2
=  \frac{4 \mu_r^2}{\pi} f_p^2 \left[Z + \frac{f_n}{f_p} (A-Z)\right]^2
\end{equation}
where $\textbf{q}$ is the momentum transfer, $\mu_r=(m_{nucl}m_{D})/(m_{nucl}+m_{D})$ and $v$ is the relative velocity. The couplings of DM to the proton and neutron, $f_p$ and $f_n $, can be expressed as 
\begin{align}
f_N={m_N \over 2 m_{D}}\left( \sum_{q=u,d,s} f^N_{Tq} {\lam_{DD qq} \over m_q} + {2\over 27} f^N_{TG} \sum_{q=c,b,t}  {\lam_{DD qq} \over m_q} \right),  \\
{\text{with}} ~~~~\quad f^N_{TG} = 1-\sum_{q=u,d,s}f^N_{Tq},
\quad (N=p,n),\nonumber
\label{fN}
\end{align}
where $m_N$ is the nucleon mass, $f^N_{Tq}$ is the form factor of the nucleon (see Table~\ref{formfactors}) and $\lam_{DD qq}$ is the effective coupling of the DM particle to a $q$-flavor quark component in the nucleon. In this model, DM-quark interaction derives from  $t$-channel exchange of the SM Higgs $h$. Thus, in the limit of zero momentum transfer, the Higgs can be integrated out and the effective coupling becomes 
\begin{equation}
\lam_{DD qq}=  {a_{hDD} m_q \over  m_h^2 v}  \,,
\label{lamxxqq}
\end{equation}
with the coupling $a_{hDD}$ given in Eq.~\ref{eq:hxx} and Eq.~\ref{eq:hHH}.
The direct detection rates in our calculation also have been evaluated using package {\tt  micrOMEGAs}. 

\begin{table}[h]
\begin{centering}
\begin{tabular}{c|c|c|c}
  \hline 
$ q$ &$ u$ & $d$ & $s$ \\
  \hline 
 $f^p_{Tq}$ & 0.0153 & 0.0191 & 0.0447 \cr
  \hline 
 $f^n_{Tq}$ & 0.0110 & 0.0273 & 0.0447 \cr
 \hline
\end{tabular} 
\caption{Form factors extracted from {\tt micrOMEGAs}.}
\label{formfactors}
\end{centering}
\end{table}

Direct searches for dark matter by the LUX~\cite{Akerib:2016vxi} and PandaX-II~\cite{Tan:2016zwf} Collaborations\footnote{We find the XENON1T~\cite{Aprile:2017iyp} results published when this work is finalized. We just comment here that the benchmarks we taken in the following also evade the new upper limits.} have recently come up with the most stringent limits on the spin-independent elastic scattering of DM off nucleons. 
 In comparing the calculated scattering cross section to the limits from the experiments, we rescale the cross section by the ratio of the predicted and observed relic density to account for the reduced flux of dark matter particles in the detectors when the relic density is undersaturated, i.e., 
$\sigma_{ scaled} = \sigma_{D-nucl} \cdot  \Omega_{DM} h^2/\Omega h^2$,
with $\Omega h^2$ being the observed relic density, we take the central value 0.1199 in our calculations.

\section{Research strategy of Electroweak phase transition }
\label{sec:PT}

The strength of electroweak phase transition can be parametrized by $v_c/T_c$, 
and generally the strong first order condition $v_c/T_c>1$ is needed to prevent the washout of baryon asymmetry~\cite{Moore:1998swa}\footnote{The gauge invariance problem needs to be keep in mind when interpreting the condition, see Ref.~\cite{Patel:2011th}.}.
 For more details on
the electroweak baryogengesis, we refer to Ref.~\cite{Morrissey:2012db}, and the interplay between the $v_c/T_c$ and the generation of baryon asymmetry in the framework of EWBG can be found in Ref.~\cite{Jiang:2015cwa}.

The effective potential at finite temperature can be used to explore the Universe cooling history.
To perform the numerical analysis of vacuum structure at finite temperature $T$,
one needs the effective potential 
\begin{equation}
V_\mathrm{eff} = V_0 + V_{CW} + V_T,
\end{equation}
with $V_0$, $V_{CW}$ and $V_T$ being the tree-level, one-loop temperature-independent
and -dependent Coleman-Weinberg potentials, respectively. The tree-level potential $V_0$ could be
obtained from Eq.~\ref{eq:tre} after the field expansion. There could be multi-step phase transition depends on the vacuum structure~\cite{Profumo:2007wc,Jiang:2015cwa,Inoue:2015pza,Blinov:2015sna}. In our case, we focus on the one-step phase transition for simplicity.  More complicate scenarios are left to future studies.
In the Landau gauge ($\xi=0$) and $\overline{MS}$ scheme, the temperature-independent one-loop Coleman-Weinberg potential 
is given by~\cite{Coleman:1973jx,Brandenberger:1984cz,
Sher:1988mj, Quiros:1999jp}:
\begin{equation}
V_{CW} = \sum_i \frac{n_i}{64\pi^2} m^4_i(h) \left(\ln \frac{m^2_i(h)} {Q^2} - C_i \right).
\end{equation}
Here, all fields coupling to the Higgs are summed, $n_i$ is the number of degrees of freedom for bosonic and fermionic fields, and renormalization-scheme-dependent constants $C_i = 1/2$ for transverse gauge bosons and $3/2$ for 
for longitudinal polarizations of gauge bosons and other particles; $m^2_i(h)$ 
are the field-dependent squared masses 
for all particles. Here, 
the counterterms have been absorbed into $V_{CW}$ implicitly.

The temperature-dependent effective potential is obtained as~\cite{Quiros:1999jp,Arnold:1992rz} 
\be
V_T = 
\frac{T^4}{2\pi^2}\left(\sum_{i=\mathrm{bosons}} n_i J_B\left[m^2_i(h)/T^2\right]-\sum_{i=\mathrm{fermions}} n_i J_F\left[m^2_i(h)/T^2\right]
\right),
\ee
with the $J$ functions defined as 
\begin{align}
J_B(x) & = \int_0^\infty dt\; t^2 \ln\left[1 - \exp\left(-\sqrt{t^2 + x}\right)\right],\label{eq:JB}\\
J_F(x) & = \int_0^\infty dt\; t^2 \ln\left[1 + \exp\left(-\sqrt{t^2 + x}\right)\right].
\end{align}

To accomplish the analysis of the phase structure as a function of $T$ analytically,
high-temperature expansions are always used:
\begin{align}
T^4 J_B\left[m^2/T^2\right] & = -\frac{\pi^4 T^4}{45}+
\frac{\pi^2}{12} T^2 m^2-\frac{\pi}{6}
T \left(m^2\right)^{3/2}-\frac{1}{32} 
m^4\ln\frac{m^2}{a_b T^2} + \mathcal{O}\left(m^2/T^2\right),\label{eq:Jb_highT}\\ 
T^4 J_F\left[m^2/T^2\right]& = \frac{7\pi^4 T^4}{360}-
\frac{\pi^2}{24} T^2 m^2-\frac{1}{32}
m^4\ln\frac{m^2}{a_f T^2} + \mathcal{O}\left(m^2/T^2\right),\label{eq:Jf_highT} 
\end{align}
where $a_b = 16 a_f = 16\pi^2 \exp(3/2 - 2\gamma_E)$. 
It should be noted that the $T^2$ terms in the above
expressions can drive symmetry restoration at high temperatures, and the
non-analytic $(m^2)^{3/2}$ term in Eq.~\ref{eq:Jb_highT} is responsible for the
barrier between the symmetric (at high temperature) and broken phases(at the critical temperature).
To avoid the  breakdown of perturbation theory induced by symmetry restoration at high temperatures,
one needs to perform a resummation of 
daisy diagrams~\cite{Parwani:1991gq} by adding finite-temperature corrections to the
boson masses in Eq.~\ref{eq:JB}: 
\be
\label{rep}
m^2 \rightarrow m^2 + m(T)^2, 
\ee
with $m(T)^2$ computed from the infrared limit of the corresponding two-point function~\cite{Carrington:1991hz} and are presented in the Appendix~\ref{app:sec}, this approach is usually called Parwani's method in literatures.

\begin{figure}[!htbp]
\begin{centering}
\includegraphics[width=.4\textwidth]{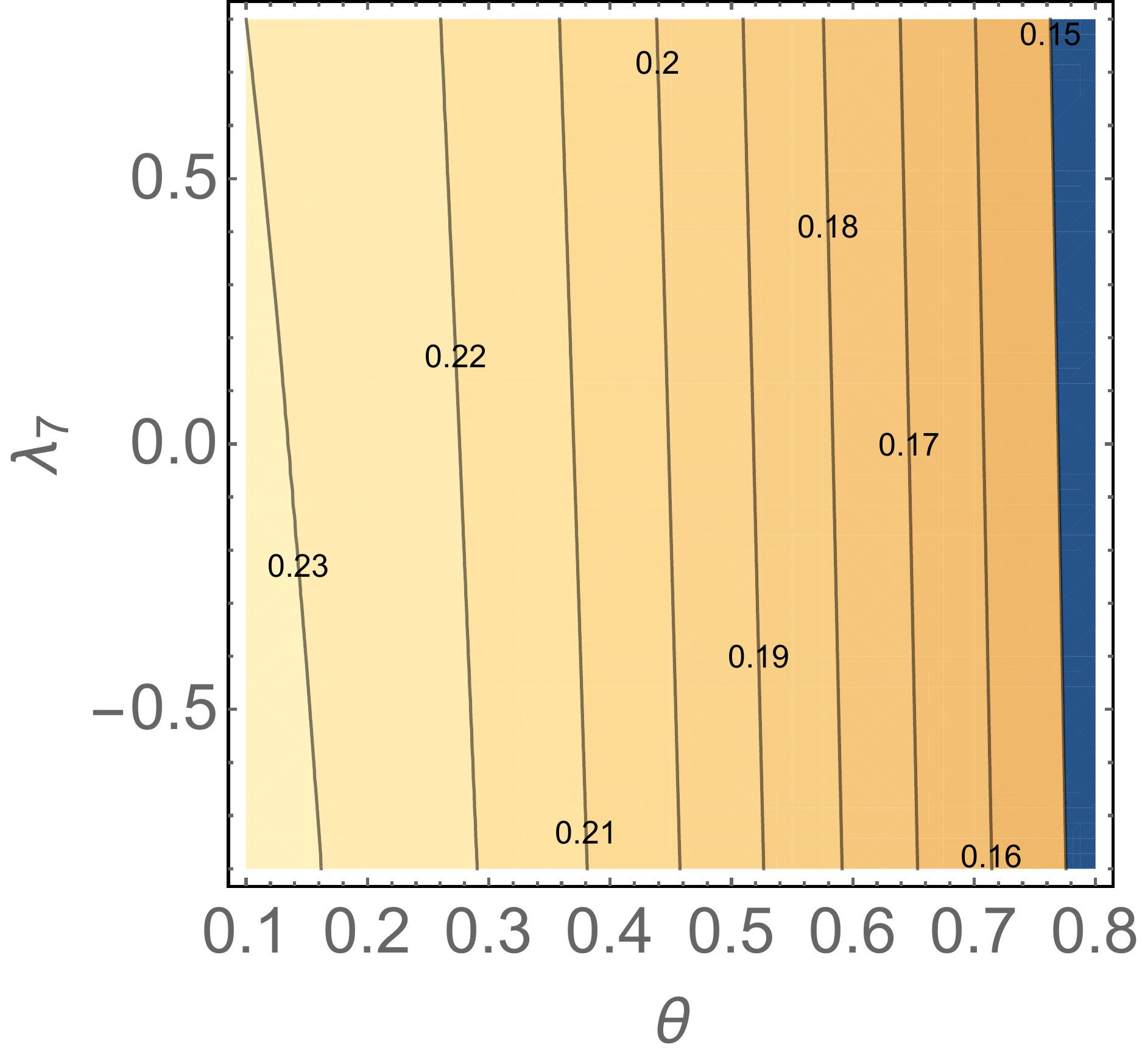}
\includegraphics[width=.4\textwidth]{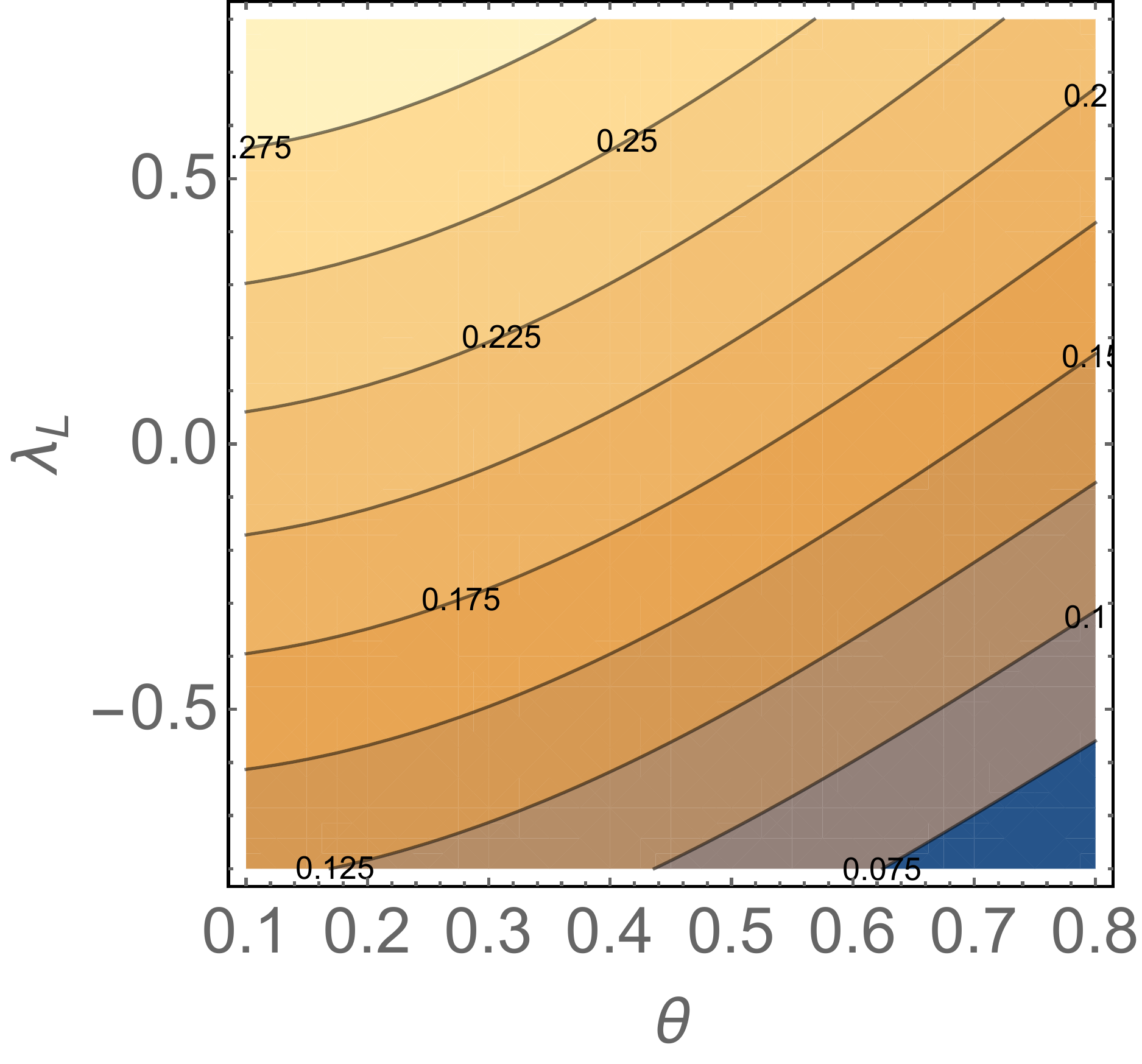}\\
\includegraphics[width=.4\textwidth]{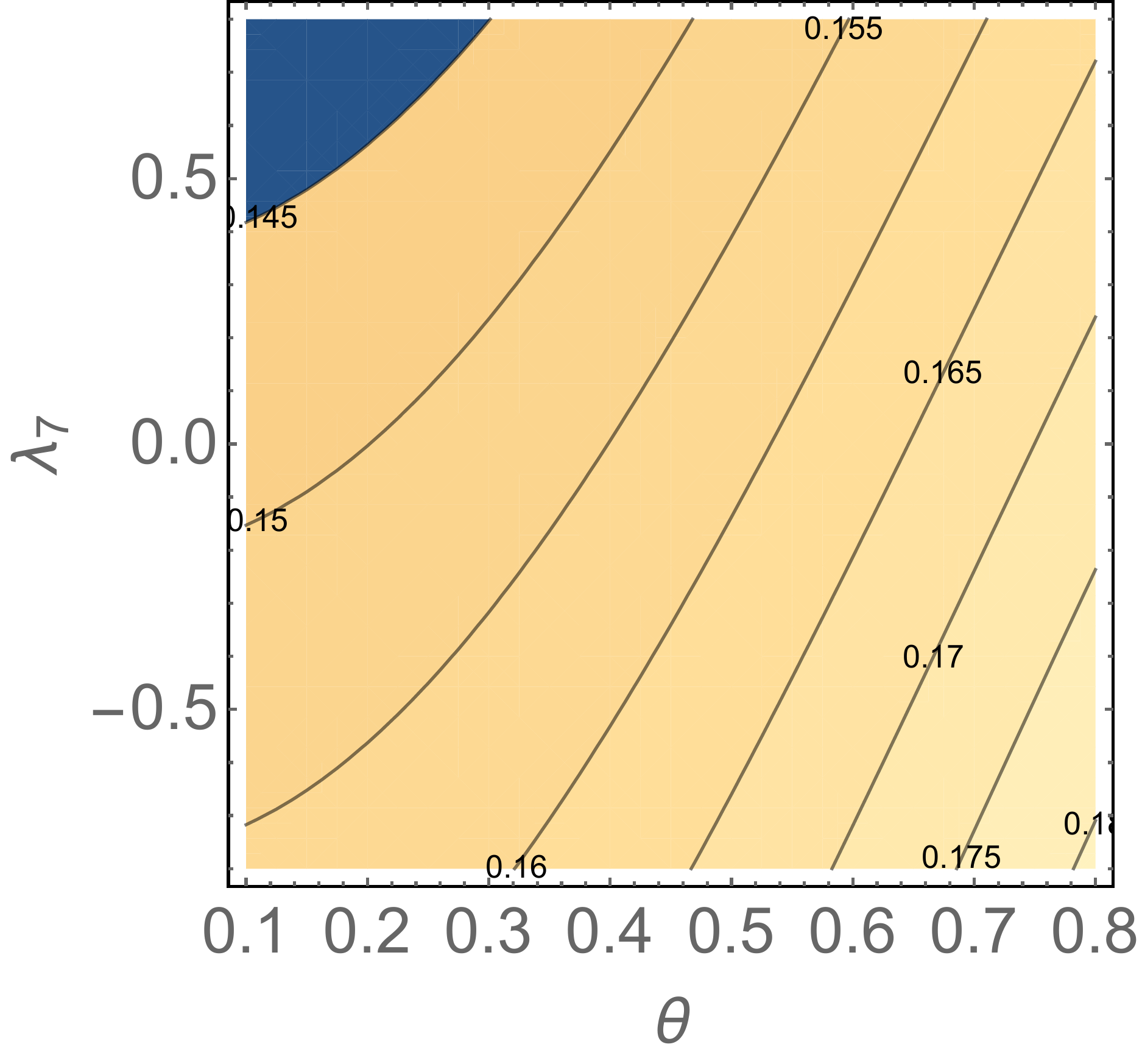}
\includegraphics[width=.4\textwidth]{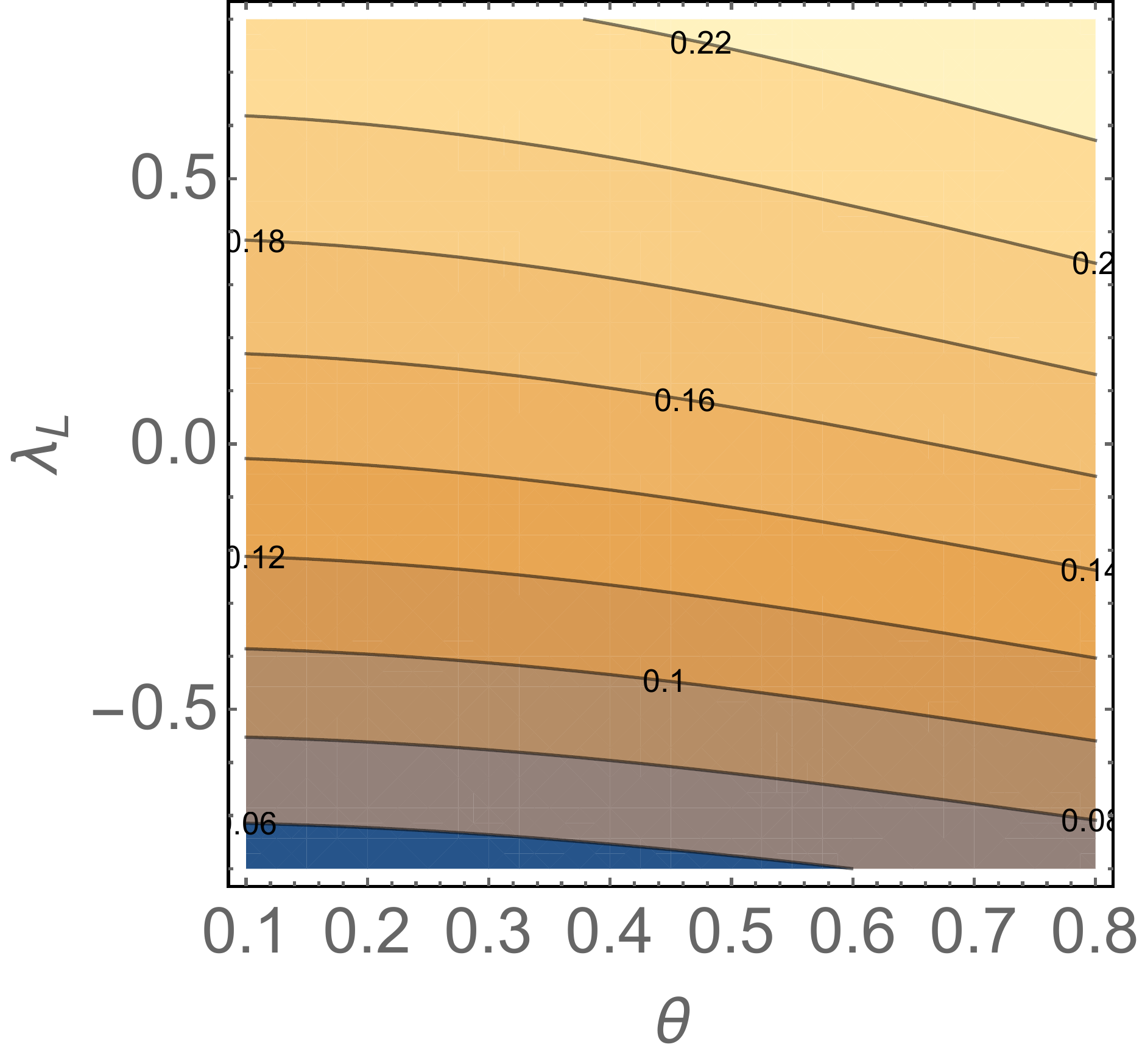}\\
\caption{The fraction $m_{SS}(T)^{3/2}/\sum_i (m^2_i)^{3/2}$ for two representative benchmarks, the top and bottom panels corresponds to the BP5  and BP-c2 scenarios which will be studied in the next section.} \label{fig:2BP_vac}
\end{centering}
\end{figure}

Here, the thermal masses contribution of $m_{SS}(T)$ gives an additional contribution to the $(m^2)^{3/2}$ term\footnote{It should be noted that the $(m^2)^{3/2}$ here means $(m^2 + m(T)^2)^{3/2}$.  } of Eq.~\ref{eq:Jb_highT}  in comparison with the IDM case. In Fig.~\ref{fig:2BP_vac} we depict the fraction  $m_{SS}(T)^{3/2}/\sum_i (m^2_i)^{3/2}$ to illustrate the contribution in various parameter spaces. We find that the ratio is more than 10\%, from which one could expect the vacuum barrier
can be altered. Therefore the mixed dark matter and the mixing of CP-even neutral scalars might affect the universe cooling process, which will be explored in the following section.

\section{DM and EWPT phenomenology}

In the low mass region of dark matter, the Ref.~\cite{Borah:2012pu,Gil:2012ya,Blinov:2015vma} found the dark matter relic abundance and the strong first order EWPT can be accomplished in the {\it Higgs funnel regime} in the IDM.  The bounds of the current direct detection on the high mass region is much weaker, which is caused by the lower number density for high mass DM. Therefore, we alternatively focus on the intermediate dark matter mass region. 
We investigate if the contributions of dark matter and the other CP-even scalar to the vacuum barrier help to realize the strong first order EWPT.

\label{sec:DMP}

\subsection{Benchmarks for DM from different mass regions } 
\label{omega}

In this subsection, we present several benchmarks for distinct DM mass regions.  
After imposing the Higgs collider bounds by using
\texttt{HiggsBounds}~\cite{Bechtle:2008jh,Bechtle:2011sb,Bechtle:2013wla},  and checking rate
measurements with
\texttt{HiggsSignals}~\cite{Bechtle:2013xfa},
all the parameter setting in each benchmark model agree with the constraints aforementioned. The quartic coupling $\lambda_2, \lambda_6, \lambda_8$ are not involved in the DM annihilation or DM-nucleus scattering processes, so we fix them only by following the constraints from the unitarity and vacuum stability conditions. 

\subsubsection{The original IDM case}
At the beginning, we briefly study the DM phenomenology in the original IDM for comparison. By setting $\lambda_6=\lambda_7=\lambda_8=\theta=0$, we restore to the IDM scenario. 
Two benchmark points are shown in Table~\ref{BP-idm}. 

\begin{table*}[!htp]
 \begin{centering}
 \caption{Benchmarks for the original IDM, here $H$ is the DM candidate.}\label{BP-idm}
  \begin{tabular}{ccccccccc} \hline\hline
   & $m_H$ & $\lambda_L$ &  $m_{A}=m_{H^{\pm}}$ & $\Omega_{DM} h^2$&$\sigma_{D-nucl}$ \\ \hline
  IDM-BP1 &~{57.5GeV}~   & ~0.001~&  ~  450GeV ~ &~  0.110 ~&~1.05$\times10^{-11}$  pb ~ ~\\ \hline
  IDM-BP2 &~{60GeV}~ & ~0.00034~&~ 450GeV ~ &~  0.103 ~&~1.12$\times10^{-12}$ pb ~\\ \hline
  \hline
  \end{tabular}
 \end{centering}
\end{table*}

Firstly, we take $m_H$ to be smaller than the gauge bosons, the main DM annihilation channel is $HH\to h\to b\bar{b}$. To avoid an overlarge cross section which leads to the relic under-abundance, we expect the DM-Higgs coupling $a_{hHH}=2 v \lambda_L$ to be small enough. As the DM mass increasing, a much smaller $\lambda_L$ is required, as could be found from IDM-BP1 and IDM-BP2 models where the DM masses are in the {\it Higgs funnel} region.  When the mass of the DM becomes larger than the gauge boson's, the annihilations to the bosons are much efficient, the DM is under-abundant even the coupling $\lambda_L \to 0$.  Then the degenerate case is the only viable scenario, for example, a parameter set can be $m_H=550$GeV, $m_A=m_{H^{\pm}}=555$GeV, and also $\lambda_L\to 0$ \footnote{In the original IDM, the intermediate mass 80-500GeV are excluded by direct detection~\cite{Goudelis:2013uca}. 
}.

\subsubsection{$\chi$ as the DM candidate} 
In the singlet-doublet scalar mixed model, we first investigate the possibility of $\chi$ being the DM candidate. We  identify two benchmarks given in Table~\ref{BP}, where the correct DM relic density can be matched and the magnitude of the cross sections of scattering off nucleus are below the LUX (or PandaX-II) strictly upper limits.
When $m_\chi<m_V$($V=W^\pm,Z$), the dominant annihilation channel is $\chi\chi\to h\to b\bar{b}$, then the DM-Higgs coupling $a_{h\chi\chi}$ characterizes the relic density as well as the DM-nucleus scattering.
For $m_\chi>m_V$ (BP2), 
the annihilations to gauge bosons  could be dominant which lead to the DM under-abundant (as discussed in the original IDM case). Due to the DM-gauge boson quartic couplings $\lambda_{\chi\chi VV}$ are proportional to $\sin^2\theta$, we need a smaller $\theta$ in order to avoid the larger cross section of $\chi\chi\to VV$. Above the $hh$ threshold with a small $\theta$, the channel $\chi\chi\to hh$ is the dominant one.

\begin{table*}[!htp]
\begin{footnotesize}
 \begin{centering}
 \caption{$\chi$ as the DM candidate, with $\lambda_2=\lambda_6=\lambda_8=0.1$.}\label{BP}
  \begin{tabular}{ccccccccc} \hline\hline
   & $m_\chi$ & $m_H$ & $\lambda_L$ & $\lambda_7$& $\theta$& $m_{A}=m_{H^{\pm}}$ & $\Omega_{DM} h^2$&$\sigma_{D-nucl}$ \\ \hline
  BP1 &~{60GeV}~  &~75GeV~ & ~0.1~  & ~-0.001 ~& ~  0.1 ~& ~  450GeV ~ &~  0.119 ~&1.01$\times10^{-12}$  pb \\ \hline
  BP2 &~{240GeV}~ &~280GeV~& ~0.045~  & ~0.06 ~& ~  0.36& ~  450GeV ~ &~  0.120 ~&2.69$\times10^{-10}$ pb \\\hline\hline
  \end{tabular}
 \end{centering}
 \end{footnotesize}
\end{table*}

\begin{table*}[!htp]
\begin{footnotesize}
 \begin{centering}
 \caption{$\chi$ as the DM candidate, with $\lambda_2=0.3, \lambda_6=\lambda_8=0.1$.}\label{BP_decay}
  \begin{tabular}{ccccccccc} \hline\hline
   & $m_\chi$ & $m_H$ & $\lambda_L$ & $\lambda_7$& $\theta$& $m_{A}=m_{H^{\pm}}$ & $\Omega_{DM} h^2$&$\sigma_{D-nucl}$ \\ \hline
  BP3 &~54GeV~ &~61GeV~& ~0.01~  & ~0.001 ~& ~ 0.1~& ~  450GeV ~ &~  0.094 ~&~1.1$\times10^{-11}$ pb ~\\ \hline
  BP4 &~240GeV~ &~370GeV~& ~0.1~  & ~0.12 ~& ~ 0.36~& ~ 450GeV ~ &~  0.101 ~&~3.77$\times10^{-10}$ pb ~\\ \hline
  BP5 &~240GeV~ &~370GeV~& ~0.12~  & ~0.115 ~& ~ 0.35~& ~ 330GeV ~ &~  0.102 ~&~2.51$\times10^{-10}$ pb ~\\ \hline\hline
  \end{tabular}
 \end{centering}
 \end{footnotesize}
\end{table*}

Three benchmarks with conditions $m_h>m_H+m_\chi$ or $m_H>m_\chi+m_h$ are listed Table~\ref{BP_decay}, where $\chi$ is the DM candidate. 
Here, considering the stability constraints, the $\lambda_2$ that does not take part in the DM phenomenology is taken to be a larger value in comparision with other benchmark scenarios in Table~\ref{BP} and Table~\ref{BP2}.
For the BP3, the Higgs invisible decay channel is open, the masses of both $H$ and $\chi$ are smaller than $m_h/2$ in comparison with the BP1, three decay channels $h\to\chi\chi, h\to H\chi, h\to HH$ should be considered. 
For the BP4 and BP5, the decay channel $H\rightarrow \chi h$ is involved in the decay chains of $e^+e^-\rightarrow HA$. We will discuss the constraints from collider experiments on parameter space and possible signals searching at colliders in the Sec.~\ref{sec:cllider}. 

\subsubsection{$H$ as the DM candidate}\label{H-DM}

For a comparison between the scenarios of $\chi$ and $H$ being the DM candidates, here we investigate the case of $m_H<m_\chi$ by employing the benchmarks in Table~\ref{BP2}.
The anatomy of dark matter phenomenology in different benchmark models are discussed in the following.

\begin{table*}[!htp]
\begin{footnotesize}
 \begin{centering}
 \caption{ $H$ as the DM candidate, with $\lambda_2=\lambda_6=\lambda_8=0.1$.}\label{BP2}
  \begin{tabular}{ccccccccc} \hline\hline
   & $m_\chi$ & $m_H$ & $\lambda_L$ & $\lambda_7$& $\theta$& $m_{A}=m_{H^{\pm}}$ & $\Omega_{DM} h^2$&$\sigma_{D-nucl}$ \\ \hline
  BP6 &~75GeV~ &~{60GeV}~& ~-0.001~  & ~0.1 ~&  0.1& ~  450GeV ~ &  0.112 & 1.01$\times10^{-12}$ pb \\ \hline
  BP7 &~280GeV~ &~{240GeV}~& ~0.06~  & ~0.03 ~&  1.1& ~  450GeV ~ & 0.117 & 2.45$\times10^{-10}$ pb \\ \hline
  BP-c1&~445GeV~ &~{440GeV}~& ~0.02~  & ~-0.07 ~&  0.6& ~  450GeV ~ &  0.102 & 1.11$\times10^{-10}$ pb \\ \hline
  BP-c2 &~480GeV~ &~{448GeV}~& ~0.05~  & ~0.07 ~&  0.58& ~  450GeV ~ &  0.0979 & 3.93$\times10^{-10}$ pb \\ \hline\hline
  \end{tabular}
 \end{centering}
 \end{footnotesize}
\end{table*}

\begin{enumerate}

\item {\bf BP6}

 We exchange the two scalars' masses in comparison with BP1. The DM-Higgs triple coupling (here it is $a_{hHH}$) plays an important role in the DM relic density estimation since the annihilations to the gauge bosons are kinematically forbidden. A simply exchange of $\lambda_L$ and $\lambda_7$ compared to BP1 would satisfy the relic abundance and the direct detection cross section limits of DM,  as indicated by Eq.~\ref{eq:scalarcouplings}, Eq.~\ref{eq:hxx} and Eq.~\ref{eq:hHH}.
 
\item {\bf BP7}

When $m_{H} > m_{V}$, the pair production of $W W (ZZ)$ in the final state of DM pair annihilation processes becomes increasingly important, because the four-point interaction through gauge couplings and s-channel process start contributing significantly. As a result the cross section for DM pair annihilating to gauge bosons becomes very large, such that the estimated thermal DM relic density may be systematically below the observed DM relic density with combinations of model parameters. 
 
The DM-gauge boson quartic couplings $\lambda_{HHVV}$ are proportional to $\cos^2\theta$, so a smaller $\theta$ drives much more efficient annihilations to the gauge bosons and therefore DM relic density is not abundant. We take a $\theta=1.1$ in the benchmark model BP7. The main component of the DM is still the singlet scalar, which is the same as the BP2 by taking the $\chi$ to be the DM candidate.

\item {\bf Degenerate case}\label{degenerate}

In the IDM scenario, when scalars $H$, $A$ and $H^{\pm}$ are nearly mass-degenerate, the co-annihilation processes become important. There is a cancellation taking place between the t/u channel contributions and the four-vertex diagram~\cite{Barbieri:2006dq,Goudelis:2013uca}. In this manner, the WIMP depletion rate can be balanced by varying the mass
splitting between DM and other scalars and the parameter $\lambda_L$  to obtain the correct mixture of transverse and longitudinal gauge bosons in the final state. These solutions are always found for small mass splittings of the odd particles, and require some tuning of the value of $\lambda_{L}$. In practice the maximal allowed mass splitting is of the order $O(10)$ GeV~\cite{Goudelis:2013uca}.

In benchmark models of BP-c1 and BP-c2, the cancellation also happens augmented by the mixing between the singlet and doublet components of the DM particle.
For benchmark BP-c2, the difference from the BP-c1 case is the odd particle $\chi$ do not degenerate with $H$. The mass splitting is much bigger which varies $\mu_{\text{soft}}$ dramatically, so the $\lambda_L$ should be tuned to adjust the DM-Higgs couping $a_{hhH}$ to be smaller enough, which is required to evade the direct detection limits.

\end{enumerate}

\subsection{Electroweak phase transition assisted by the mixed dark matter}
\label{sec:EWPTDM}

We exam all the benchmarks given above to find which can yield a strong first order EWPT.
There are two benchmark models left which are summarized in Table~\ref{BP_ewpt}  along with the corresponding critical temperatures and the strengths of the phase transition.
In the following we present the combined results of both DM and EWPT by varying sensitive parameters near the benchmark points BP5 and BP-c2. 
The $\chi-H$ mixing effects are shown in Fig.~\ref{fig:resBP8_1} and Fig.~\ref{fig:results-BPcan2-1}, and other parameters effects are presented in Fig.~\ref{fig:resBP8_2} and Fig.~\ref{fig:results-BPcan2-2}. 

\begin{table*}[!htp]
 \begin{centering}
 \caption{Benchmark models in which the EWPT can be strongly first order.}\label{BP_ewpt}
  \begin{tabular}{cccccc} \hline\hline
   Benchmarks & $v_c$  & $T_c$  & $v_c/T_c$ & $\Omega_{DM} h^2$&$\sigma_{D-nucl}$ \\ \hline
   BP5 &~242.851GeV~ &~133.750GeV~ & ~ 1.8157  &~  0.102    ~&~2.51$\times10^{-10}$ pb ~\\ \hline\hline
   BP-c2 &~319.011GeV ~&~195.652GeV~ & ~1.6305~  & 0.0979      ~&~3.93$\times10^{-10}$ pb ~\\ \hline
  \end{tabular}
 \end{centering}
\end{table*}

\begin{figure}[!htp]
\begin{centering}
\includegraphics[width=.4\textwidth]{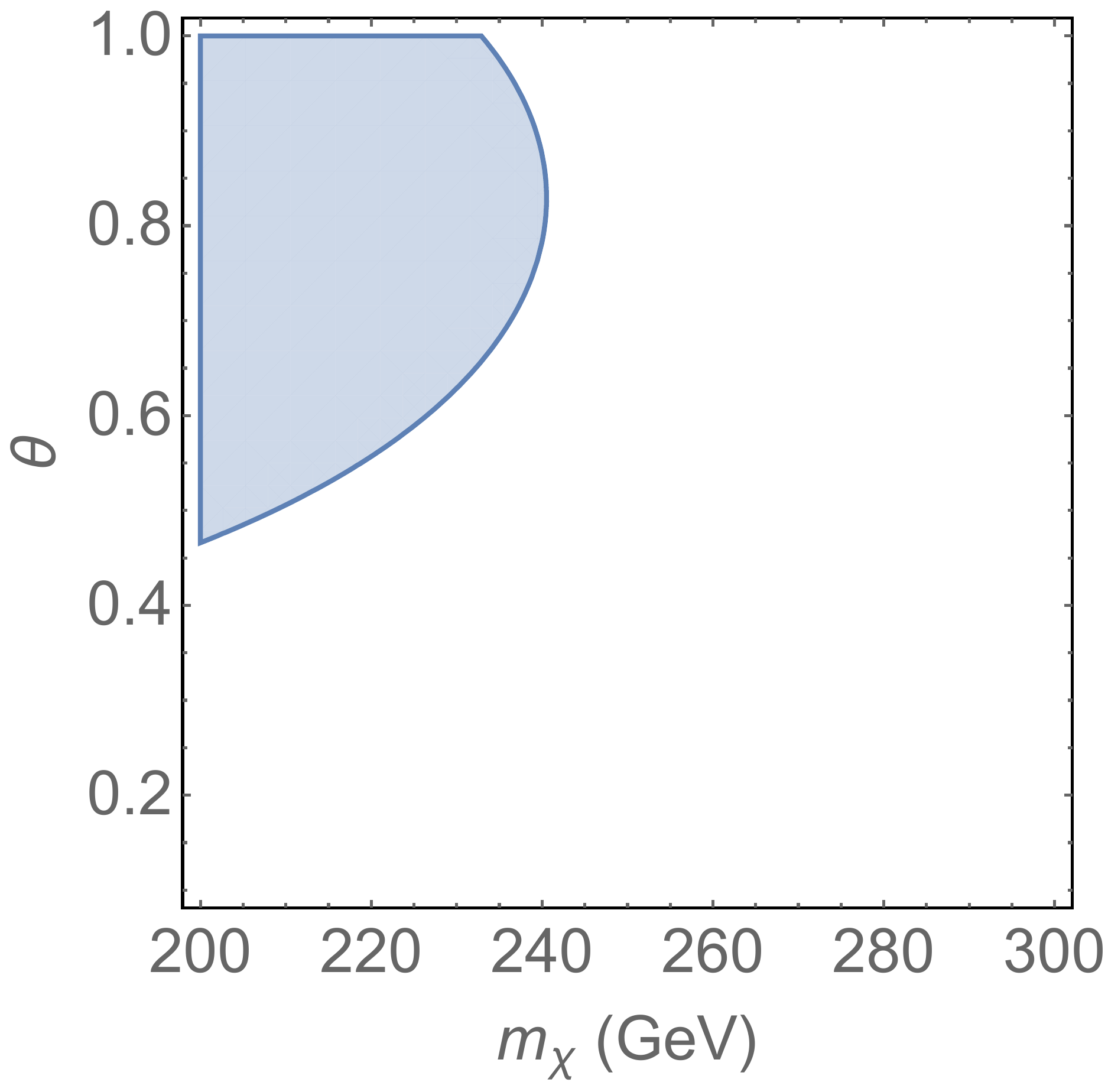}
\includegraphics[width=.4\textwidth]{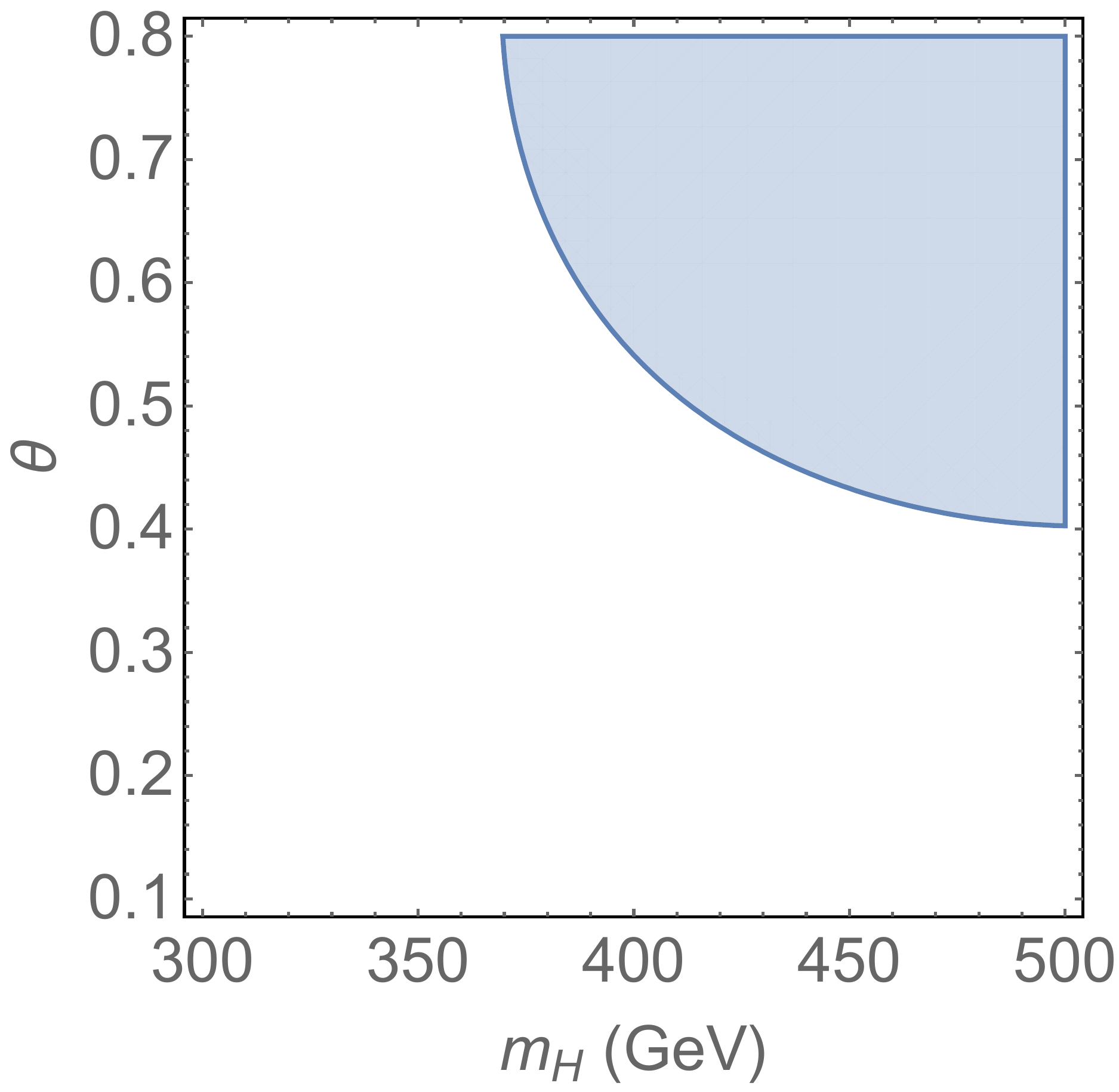}\\
\bigskip
\includegraphics[width=.4\textwidth]{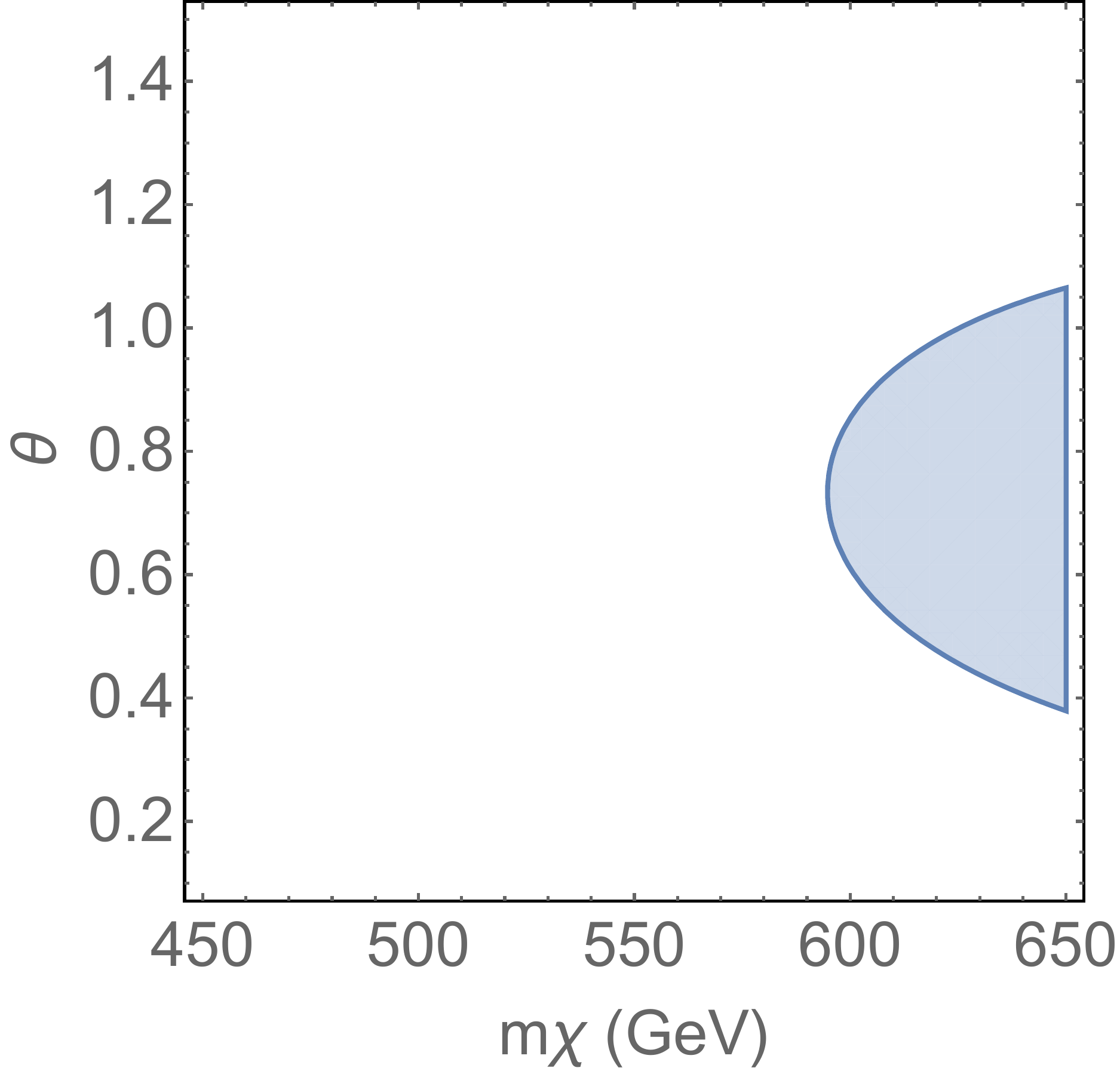}
\includegraphics[width=.4\textwidth]{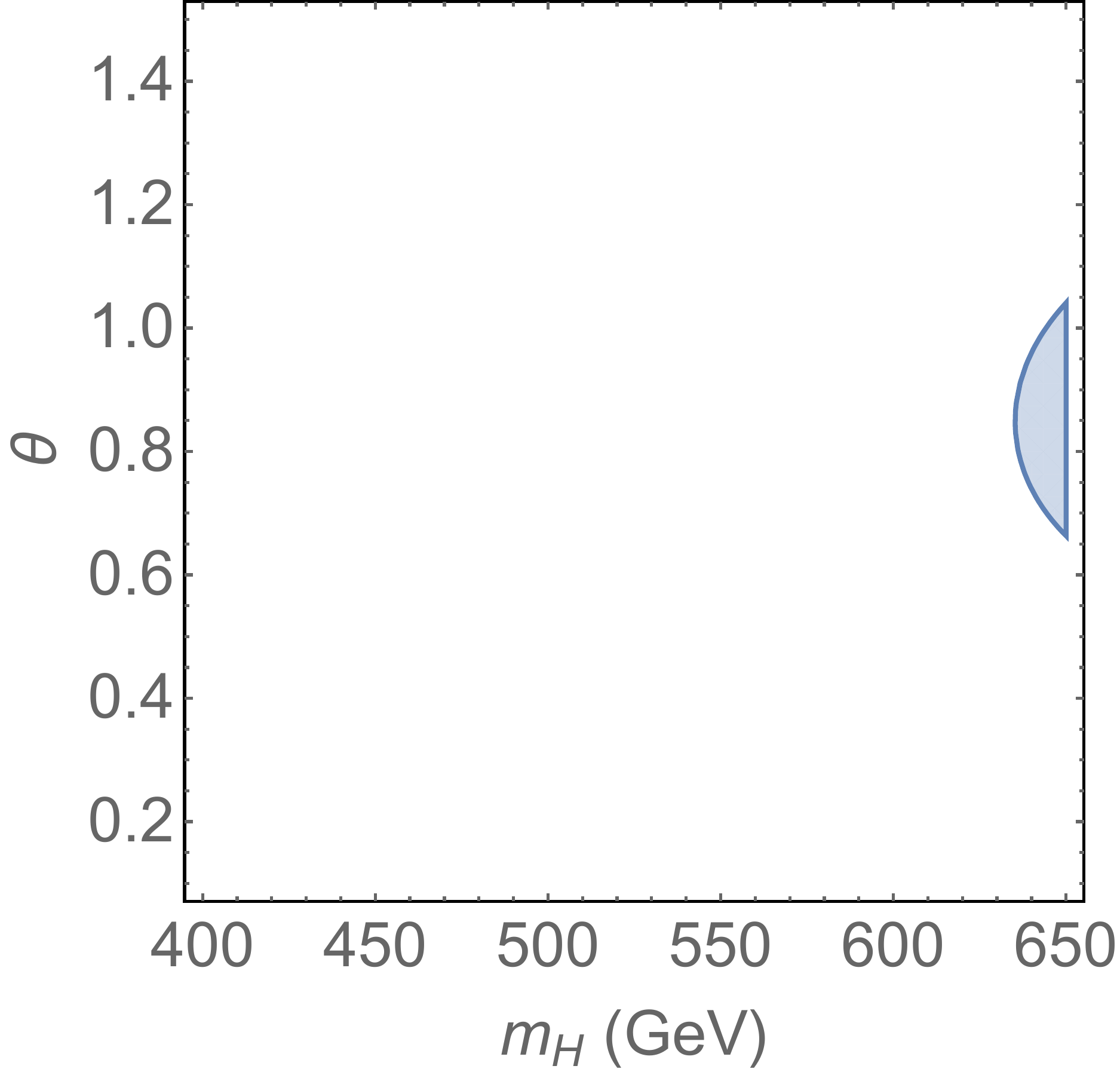}
\caption{95\% Confidence Level excluded regions from fit to the $T$ parameter for BP5(Top) and BP-c2(Bottom) benchmark scenarios. } 
\label{fig:T}
\end{centering}
\end{figure}

Before devoting to the numerical analysis of DM and EWPT, we firstly consider the electroweak precision test constraints. The $T$ parameter exclusion of two benchmark scenarios 
are shown in Fig.~\ref{fig:T}. The two plots in top panel imply that the mixing angle should be small for the DM mass around 200 GeV. For BP-c2, parameter space is less constrained by the $T$ parameter. We also perform the constraints from the stability conditions, the blue region in Fig.~\ref{fig:resBP8_1}, Fig.~\ref{fig:resBP8_2},  Fig.~\ref{fig:results-BPcan2-1}, and Fig.~\ref{fig:results-BPcan2-2} 
are excluded. One can see that the parameter space is constrained severely. 
Moreover, a large variation of the quartic coupling $\lambda_2, \lambda_6, \lambda_8$  only  leads to tinny affect on the one-step phase transition case as numerically checked by us. Therefore we focus on
the survey of the phase transition in the parameter spaces of $\lambda_L,\lambda_7,\theta,m_{H,\chi}$ with others being fixed.

\subsubsection{$\chi$-DM benchmark model: BP5}

\begin{figure}[!htbp]
\begin{centering}
\includegraphics[width=.4\textwidth]{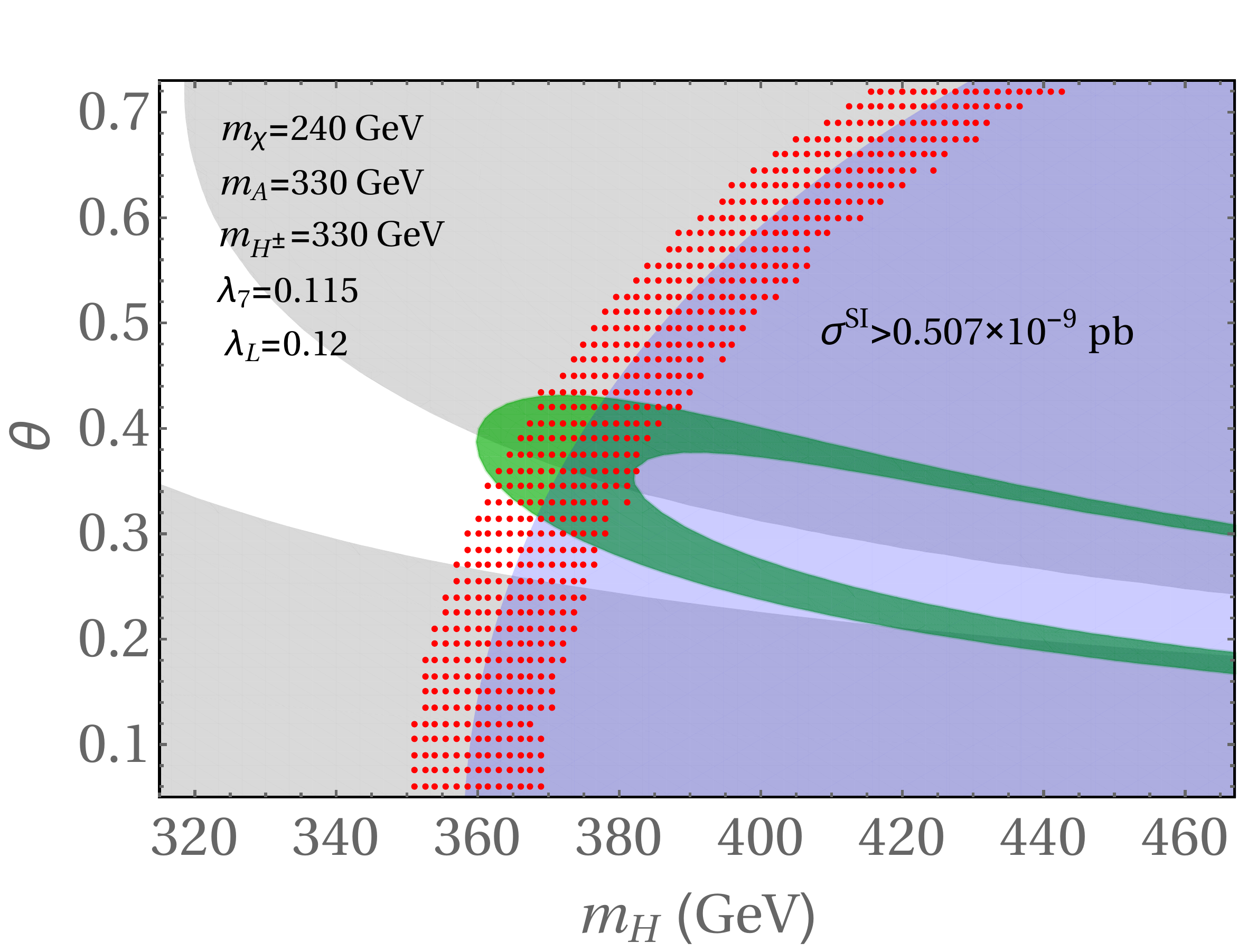}
\includegraphics[width=.4\textwidth]{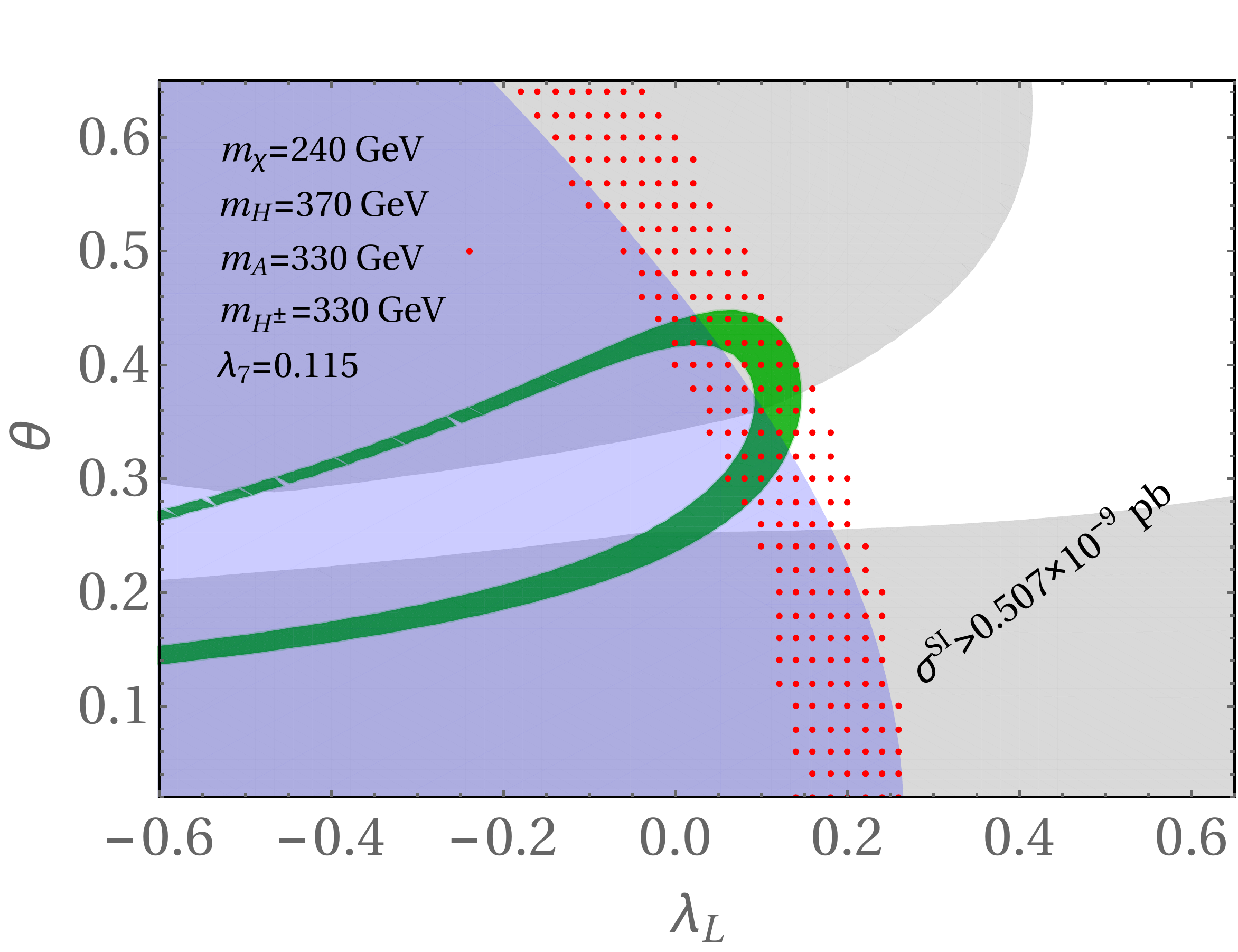}\\
\includegraphics[width=.4\textwidth]{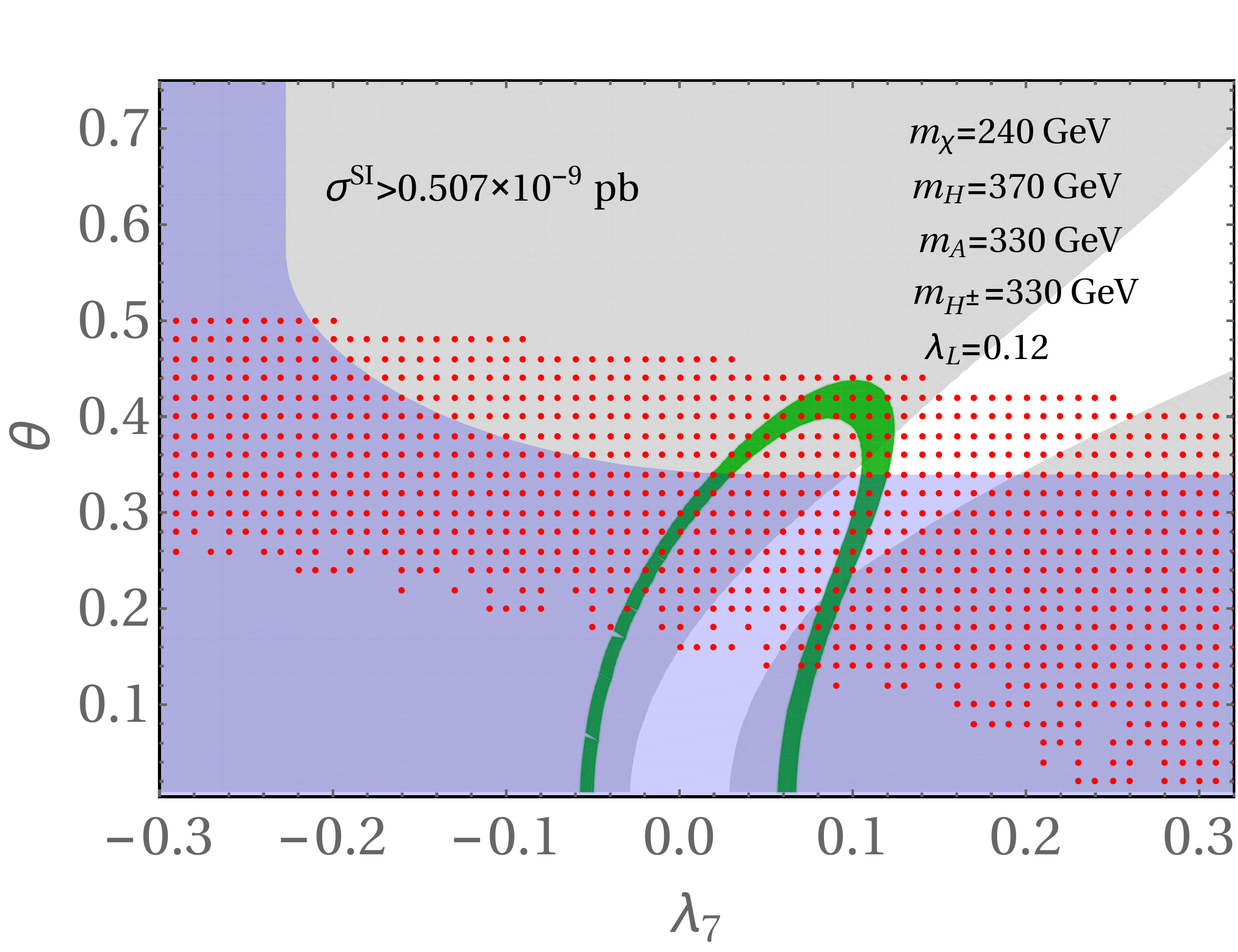}
\includegraphics[width=.4\textwidth]{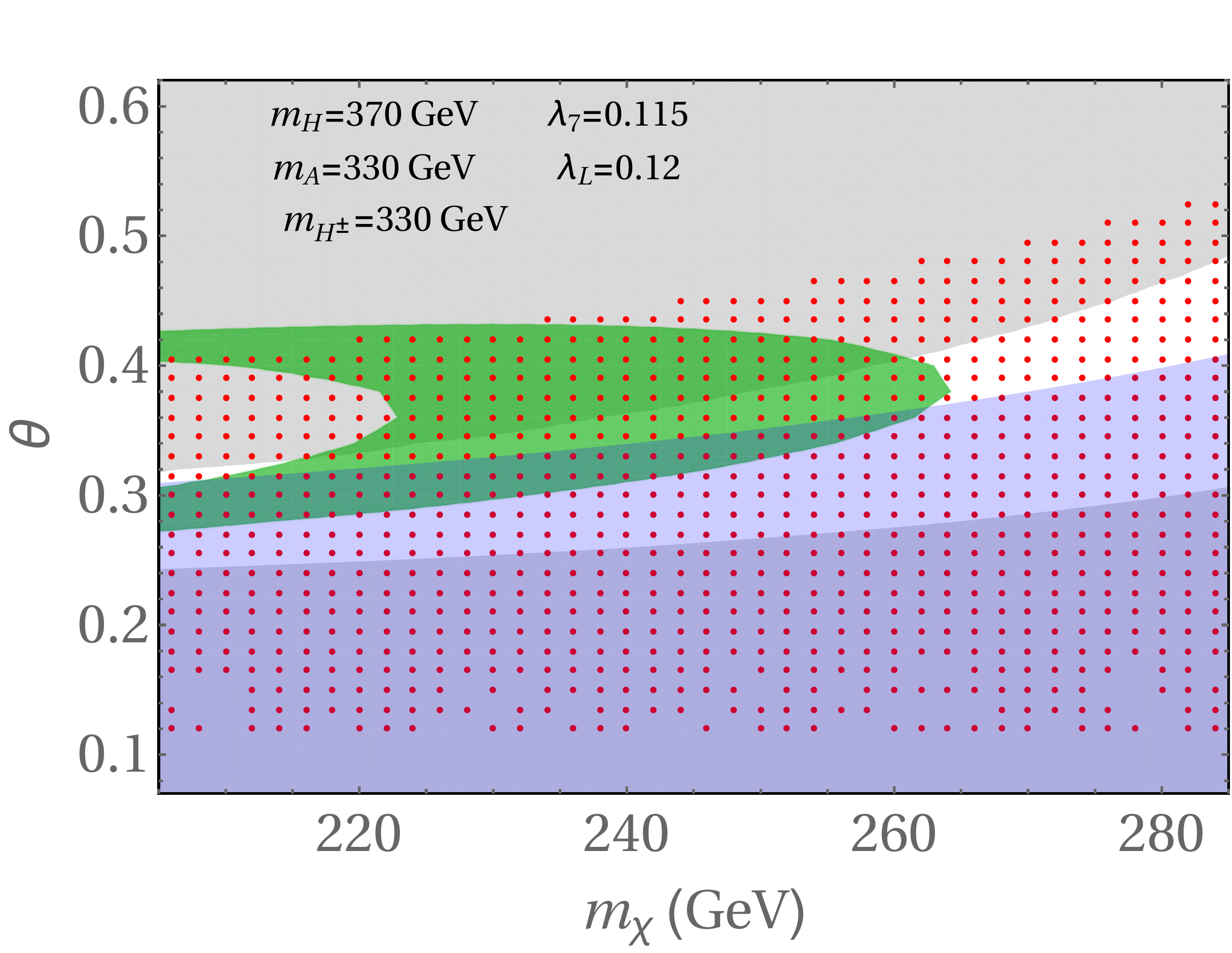}
\caption{Combined results corresponding to BP5 by varying mixing angle $\theta$ and other parameters. 
The red dotted regions are the parameter space where strong first order EWPT occurs,  
the green regions stand for $0.09<\Omega h^2 < 0.12$.  The gray regions are excluded by LUX limits. The blue shaded regions are excluded by the vacuum stability conditions.} \label{fig:resBP8_1}
\end{centering}
\end{figure}

\begin{figure}[!htbp]
\begin{centering}
\includegraphics[width=.4\textwidth]{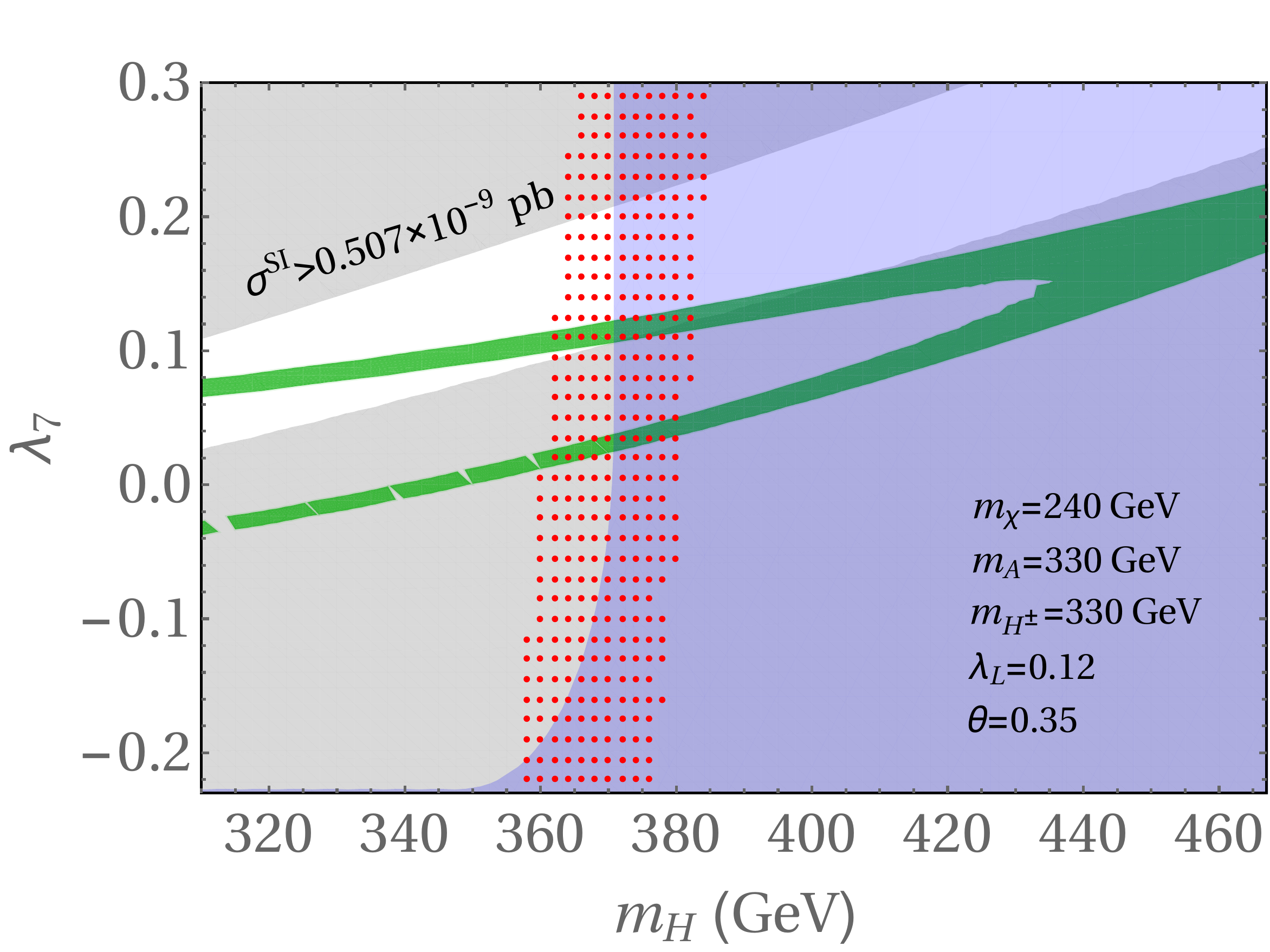}
\includegraphics[width=.4\textwidth]{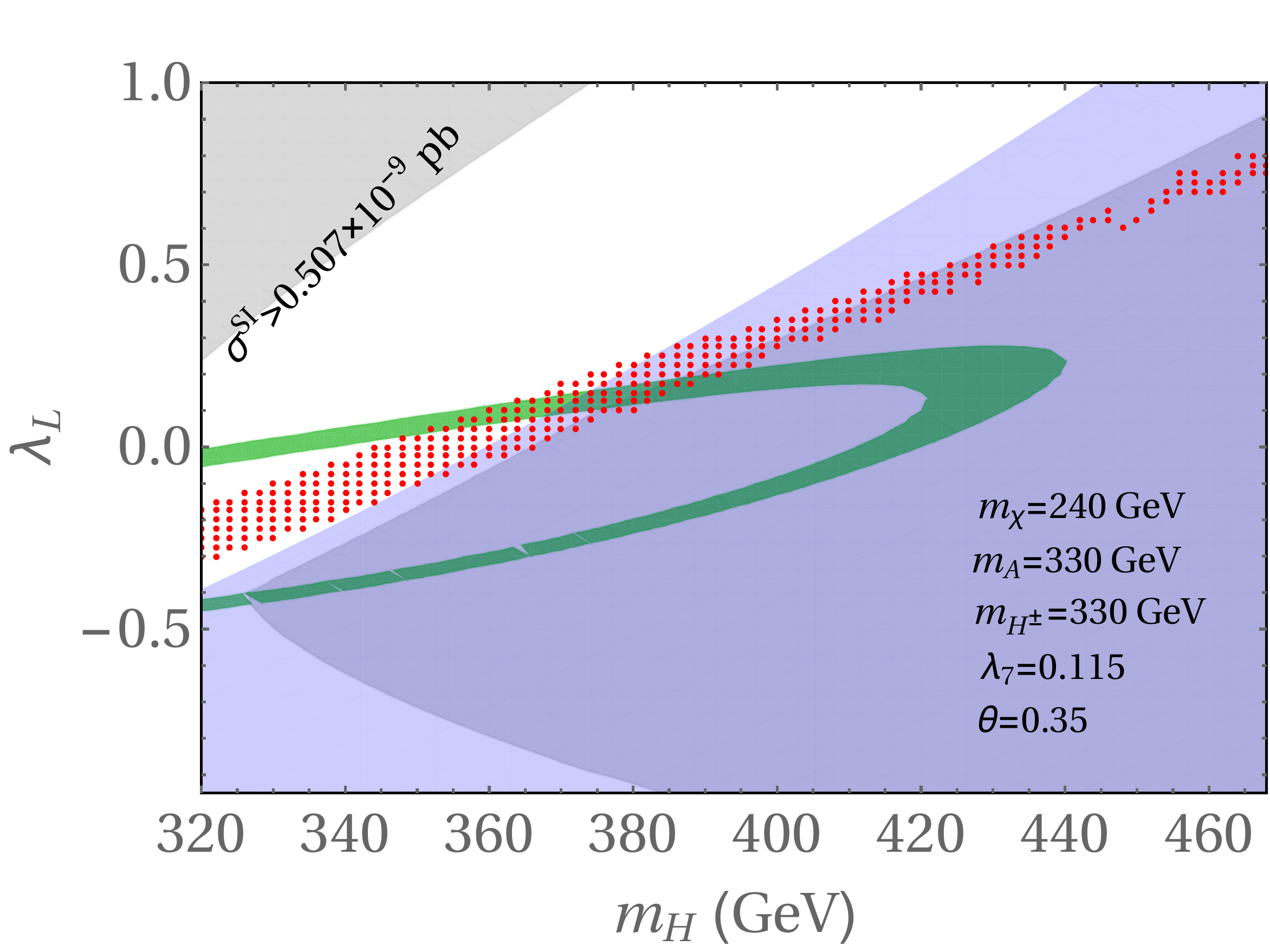}\\
\includegraphics[width=.4\textwidth]{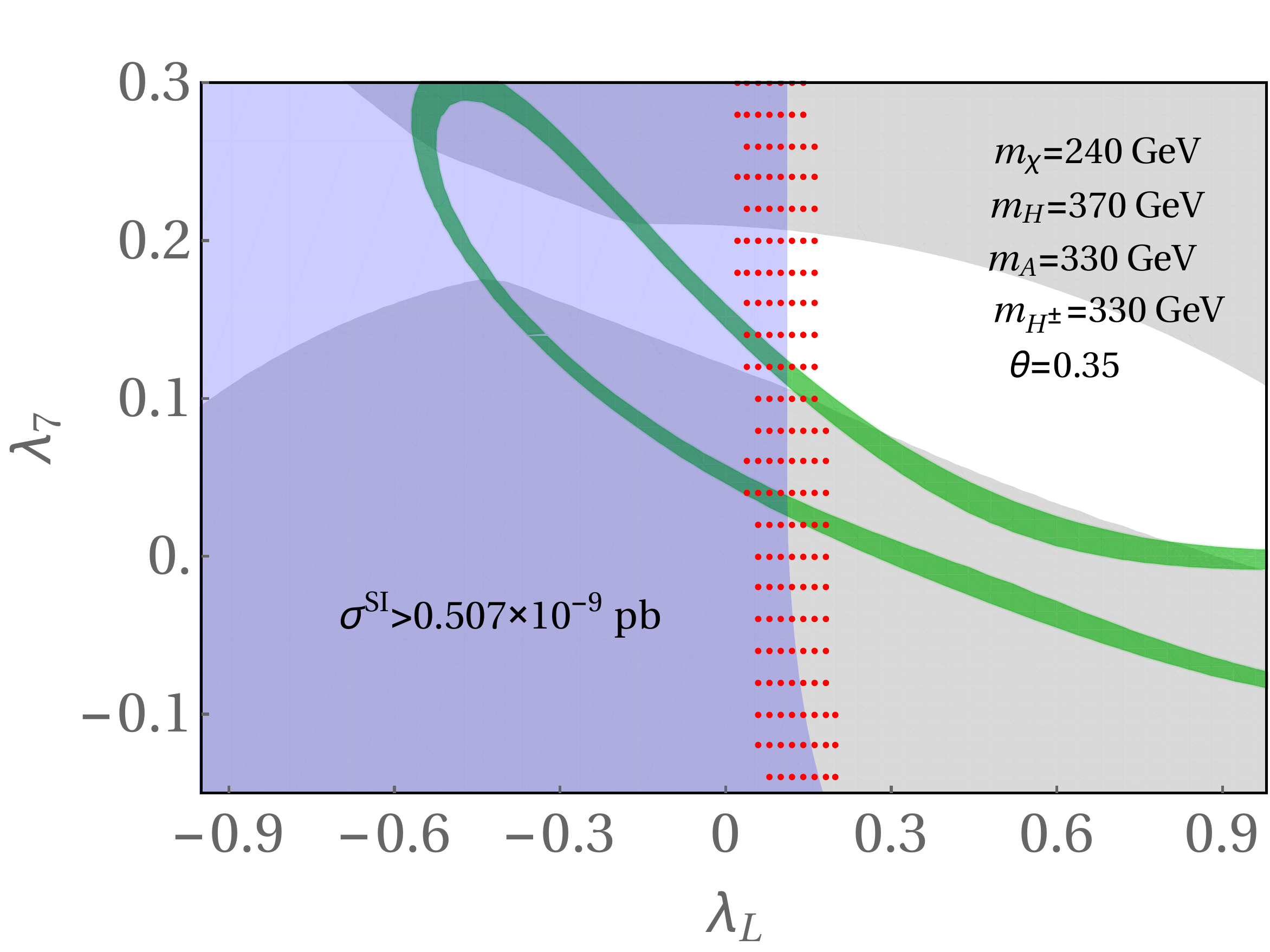}
\includegraphics[width=.4\textwidth]{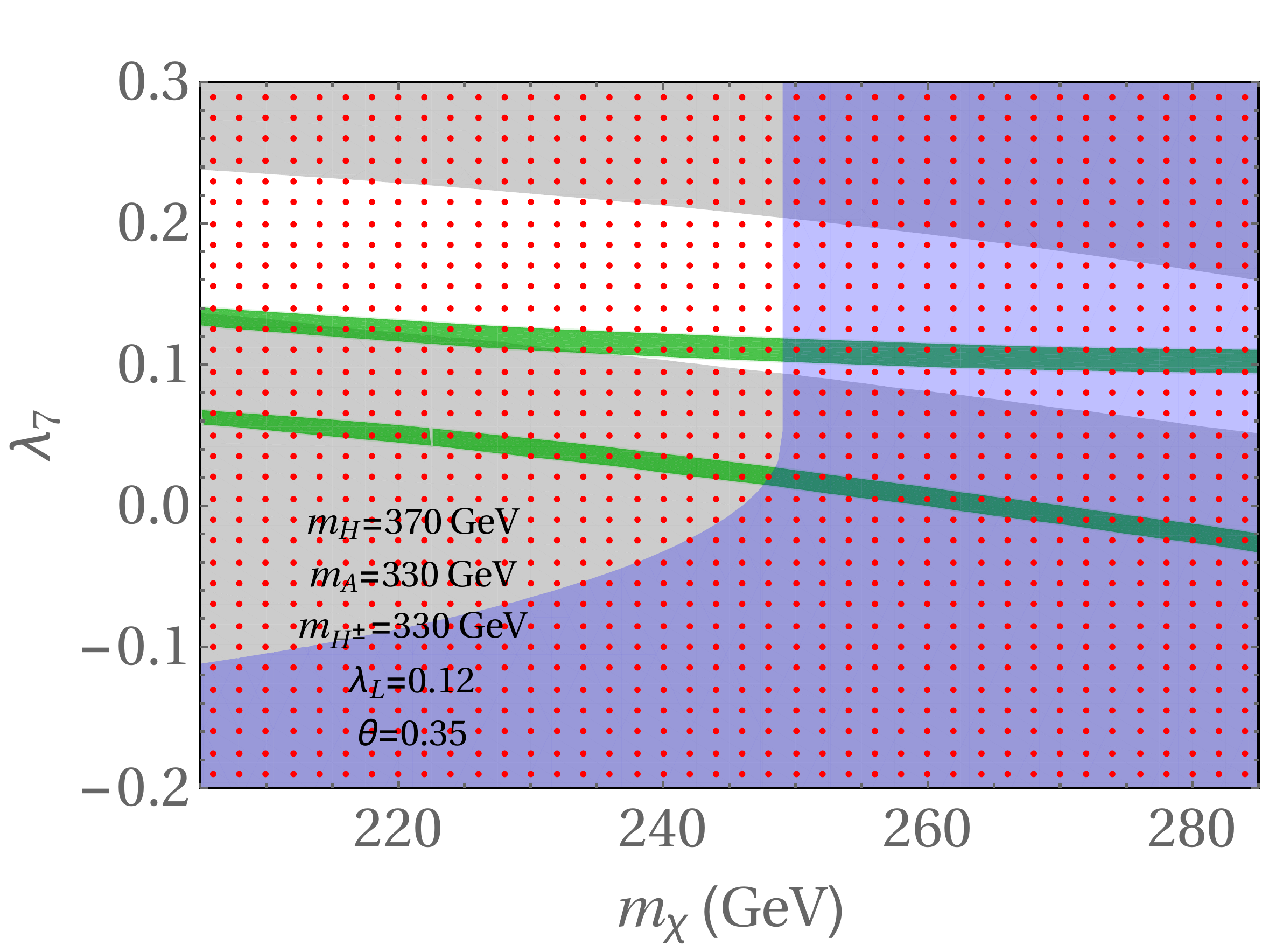}
\caption{For BP5 with fixed mixing angle $\theta$. The convention of the colors in the plots is the same as in Fig.~\ref{fig:resBP8_1}.
} \label{fig:resBP8_2}
\end{centering}
\end{figure}

We first analysis the combined results of Benchmark BP5, which are presented in the Fig.~\ref{fig:resBP8_1} and Fig.~\ref{fig:resBP8_2}.

\begin{enumerate}
\item {\bf EWPT}

As aforementioned in Sec.~\ref{sec:PT}, the vacuum barrier can be affected by the DM mass slightly. We find that the DM mass 
dependence of the strength of EWPT is tinny in this benchmark as explored in Ref.~\cite{Jiang:2015cwa}, and the whole mass range can provide a strongly first-order EWPT (red-dotted region), as depicted in the parameter spaces of $m_{\chi}-\theta$ (bottom-right of Fig.~\ref{fig:resBP8_1}) and $m_{\chi}-\lambda_7$ (bottom-right of Fig.~\ref{fig:resBP8_2}).
We observe that the strength of the EWPT subtly depends on the mass splitting $\Delta M=|m_H-m_{A,H^{\pm}}|$ and the mixing angle, and the larger $\theta$ leads to a strong first order EWPT when the mass splitting $\Delta M$ is large.
As depicted in the top-left panel of Fig.~\ref{fig:resBP8_1} and top panels of Fig.~\ref{fig:resBP8_2}, the realization of the strongly first-order EWPT prefers a narrow range of $m_H$   with the fixed values of $\theta, \lambda_7$.  
Furthermore,  a limited range of the coupling $\lambda_L$ is required by the condition $v_c/T>1$ as shown in top-right panel of  Fig.~\ref{fig:resBP8_1} and bottom-left panel of Fig.~\ref{fig:resBP8_2}.

\item{\bf DM relic density}

In this benchmark model, the scalar $\chi$ plays the role of DM. As demonstrated in the $\lambda_L-\theta$, $\lambda_7-\theta$, $m_\chi-\theta$ planes of Fig.~\ref{fig:resBP8_1}, the mixing angle $\theta$ induces efficient change of the magnitude of the DM relic density, this is because the main annihilation channel is $\chi\chi\to hh$ with a $\theta$-dependent four-vertex coupling $2\lambda_7 \cos^2\theta +2\lambda_L \sin^2\theta$. We find that the DM relic requires mixing angle to be smaller than 0.5, and then we notice that the main contribution comes from  $2\lambda_7 \cos^2\theta$ part. Thus the DM relic density sensitively depends on the parameters $\lambda_7$.

In the top-left panel of Fig.~\ref{fig:resBP8_1} and top panels of Fig.~\ref{fig:resBP8_2}, when $m_H$ becomes larger than about $400$ GeV, the main annihilation channel changes to be $\chi\chi\to t\bar{t}$ which is proportional to the DM-Higgs coupling $a_{h\chi\chi}$.  Due to the coupling $a_{h\chi\chi}$ can be affected a lot by the $m_H$ since which is involved in the parameter $\mu_{\text{soft}}$, therefore, one can expect the predicted DM relic abundance is affected by the variation of $m_H$. In the top-left panel of Fig.~\ref{fig:resBP8_1} (top panels of Fig.~\ref{fig:resBP8_2}), with the increasing of 
$m_H$ one gets an increasing (decreasing) $a_{h\chi\chi}$, and therefore a narrower (broader)  viable region for  the correct DM relic density.

\item{\bf DM direct detection}

Considering the current 
restrictive limit from LUX (and PandaX-II), we study the constraints on the parameter space of the model. The gray regions in the figures are excluded by the DM direct detection limits.

In the intermediate mass region (around 200 GeV),  we find that LUX limits constrain the parameter $\theta$ and $\lambda_7$ rigorously, due to the DM-Higgs coupling $a_{h\chi\chi}\sim 2v\lambda_7\cos^2\theta$. 
Only a small region evades the constraint in each plane of Fig.~\ref{fig:resBP8_1} and Fig.~\ref{fig:resBP8_2}.  The narrow allowed parameter spaces are due to the cancellation effects in the DM-Higgs coupling $a_{h\chi\chi}$, as can be seen from Eq.~\ref{eq:hxx} when $m_\chi< m_H$.
In the bottom-right plots of Fig.~\ref{fig:resBP8_1} and Fig.~\ref{fig:resBP8_2}, we preform the exclusion of $\theta, \lambda_7$ by varying the DM mass.
Although the limit is much strict, there is still room for obtaining the correct DM relic density and the strong first order EWPT. 

\end{enumerate}

\subsubsection{$H$-DM benchmark model: BP-c2}
\begin{figure}[!htbp]
\begin{centering}
\includegraphics[width=.4\textwidth]{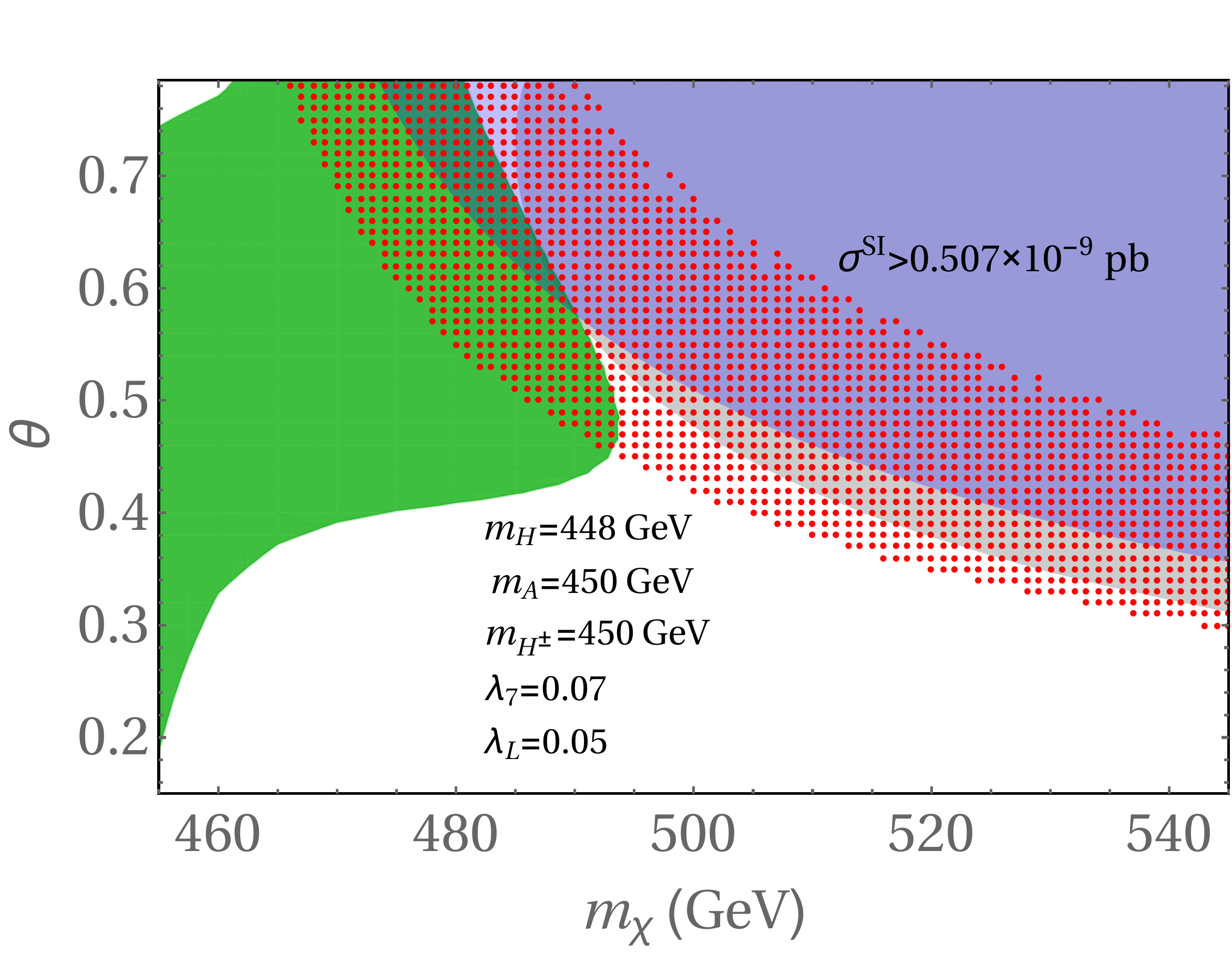}
\includegraphics[width=.4\textwidth]{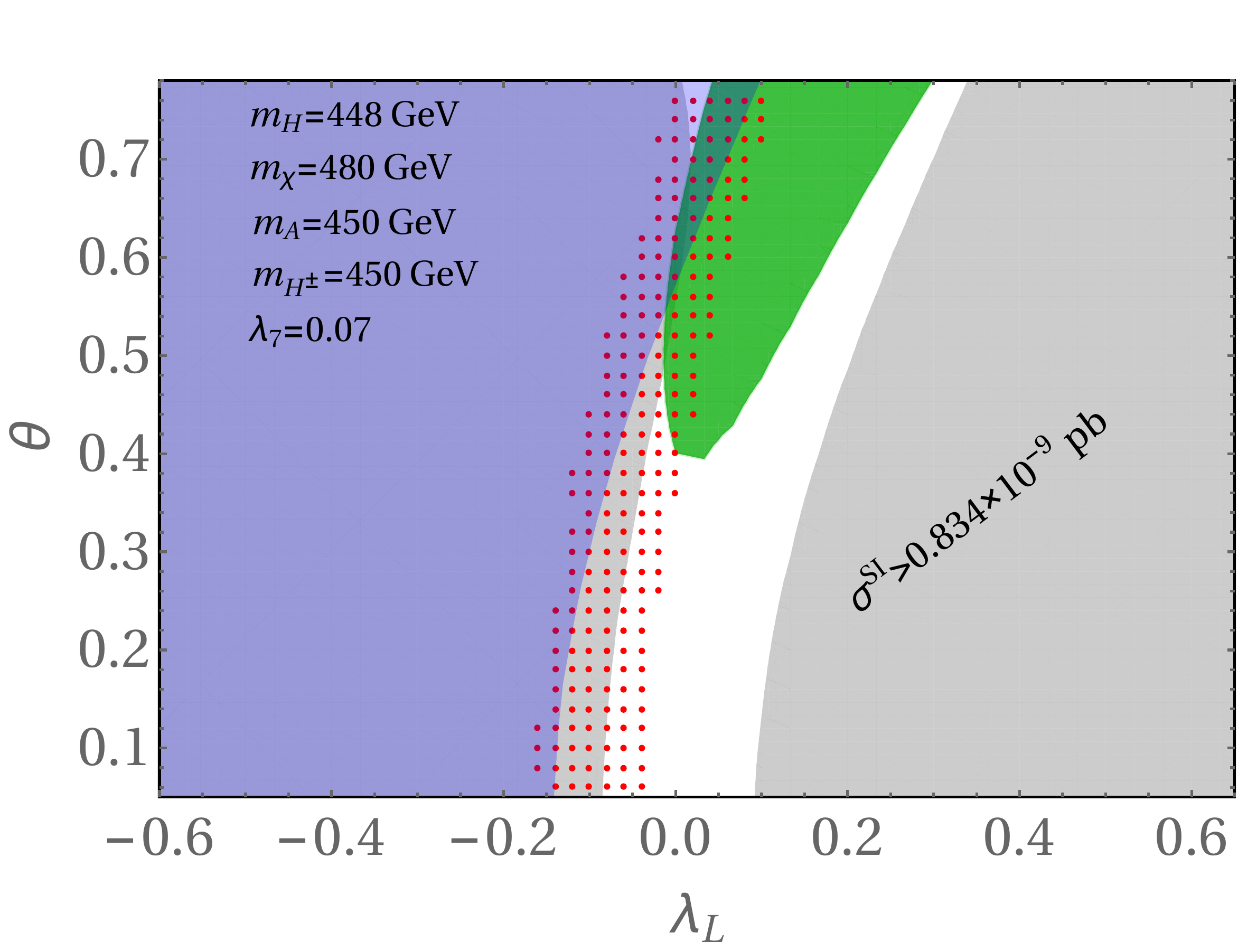}\\
\includegraphics[width=.4\textwidth]{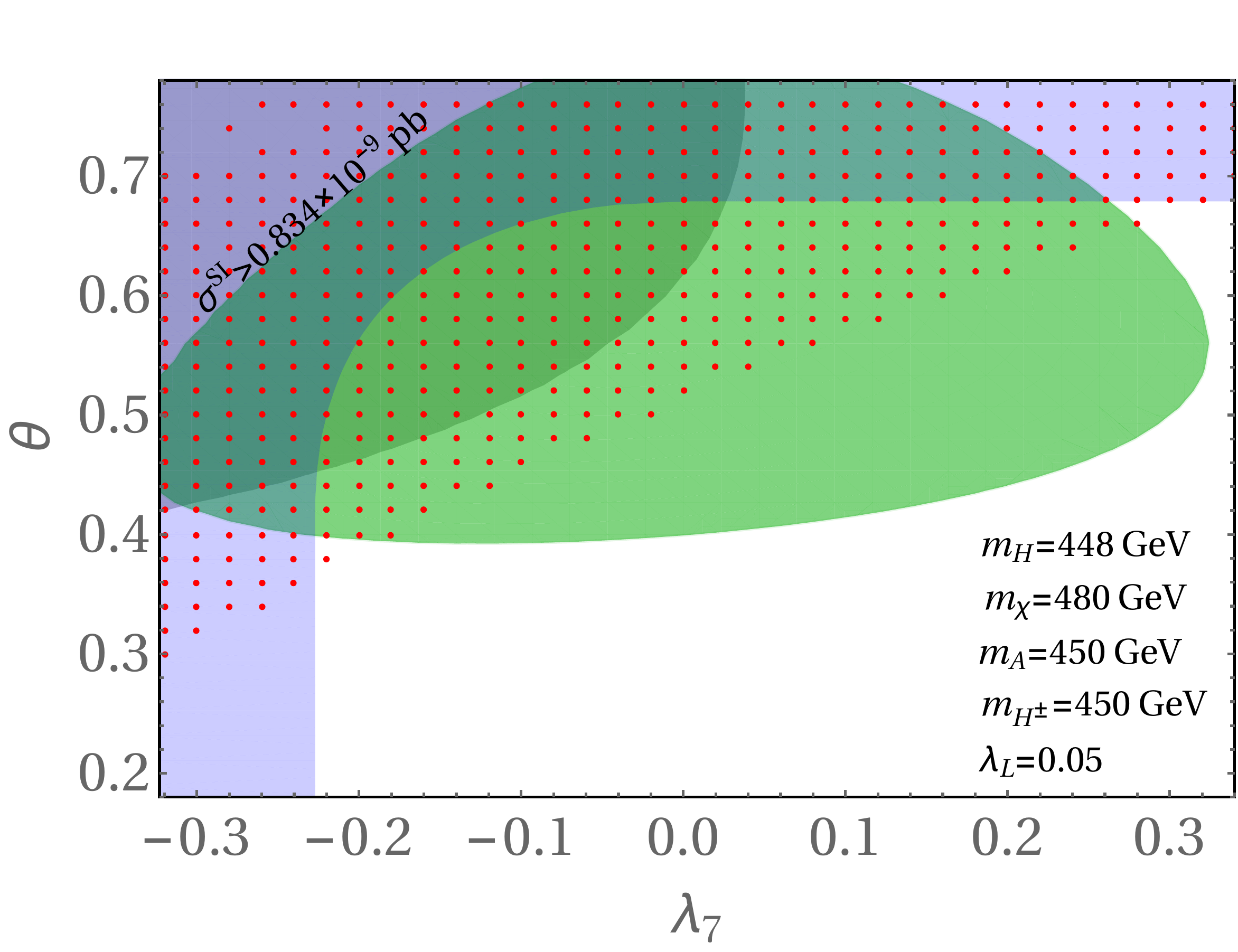}
\includegraphics[width=.4\textwidth]{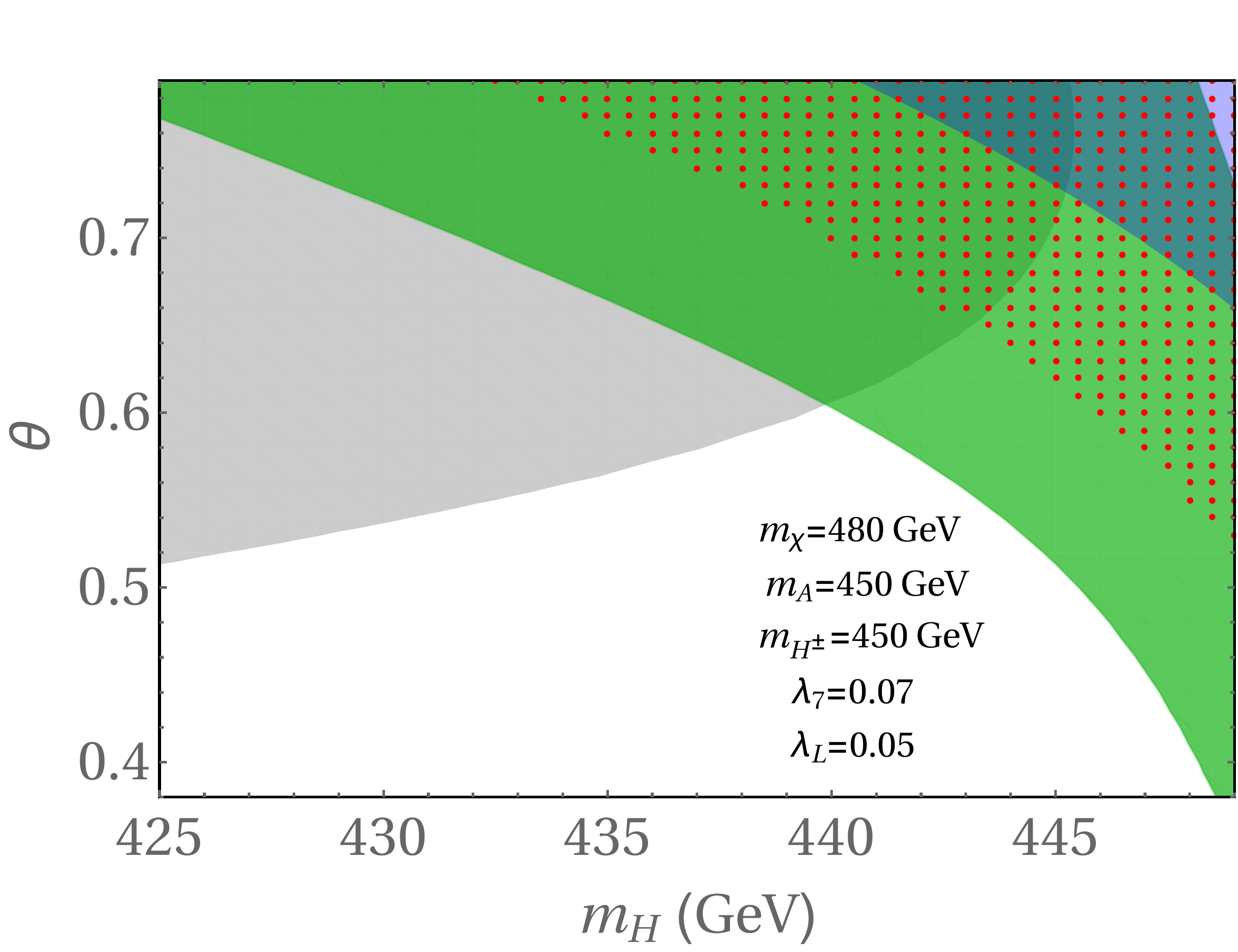}
\caption{For BP-c2 with varing $\theta$,
the green regions stand for $0.09<\Omega h^2 < 0.12$. The red dotted regions stand for the parameter space can satisfy the condition $v_c/T_c>1$. The gray regions are excluded by LUX.  The blue shaded regions are excluded by the vacuum stability conditions. } 
\label{fig:results-BPcan2-1}
\end{centering}
\end{figure}

\begin{figure}[!htbp]
\begin{centering}
\includegraphics[width=.4\textwidth]{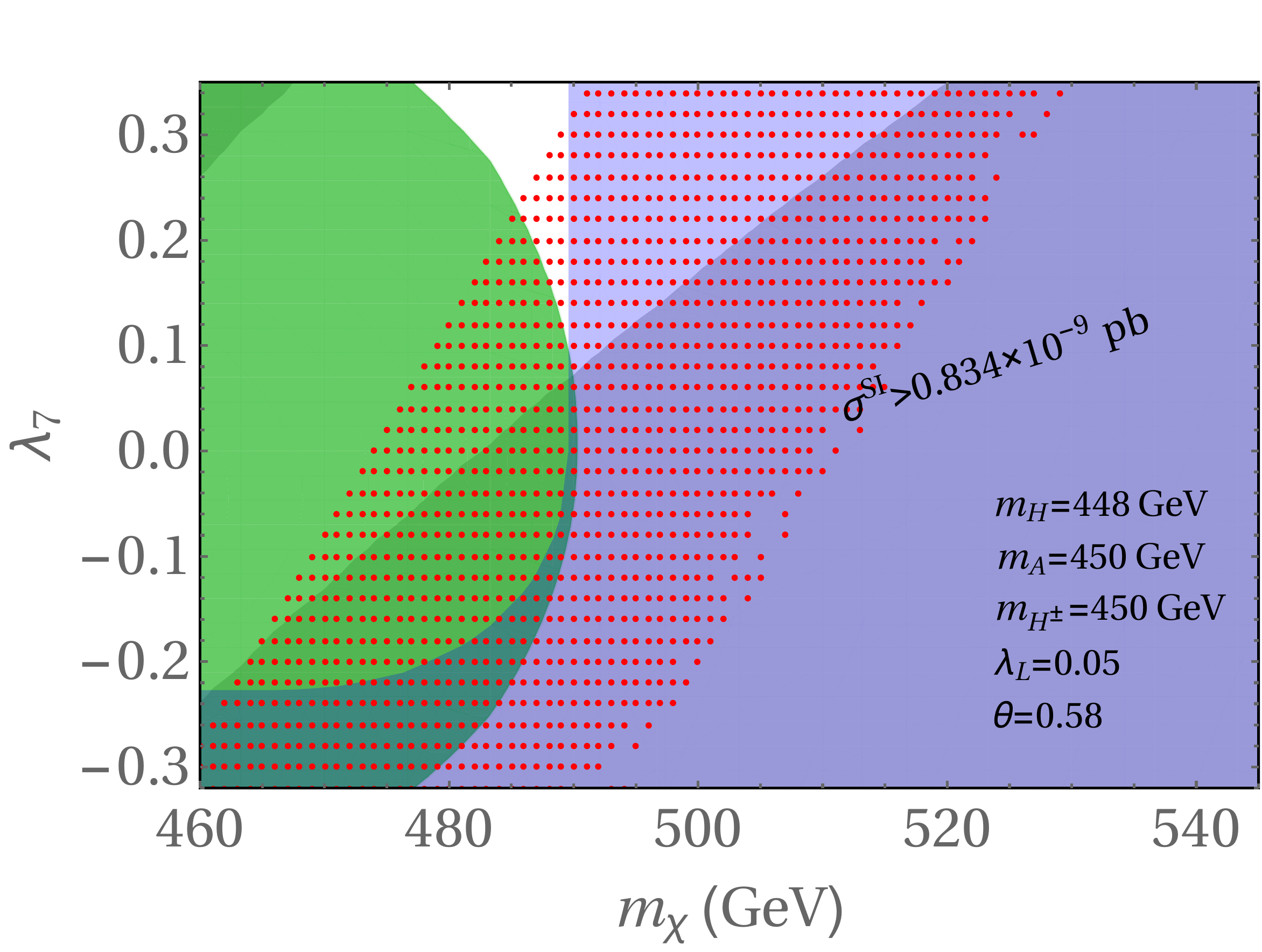}
\includegraphics[width=.4\textwidth]{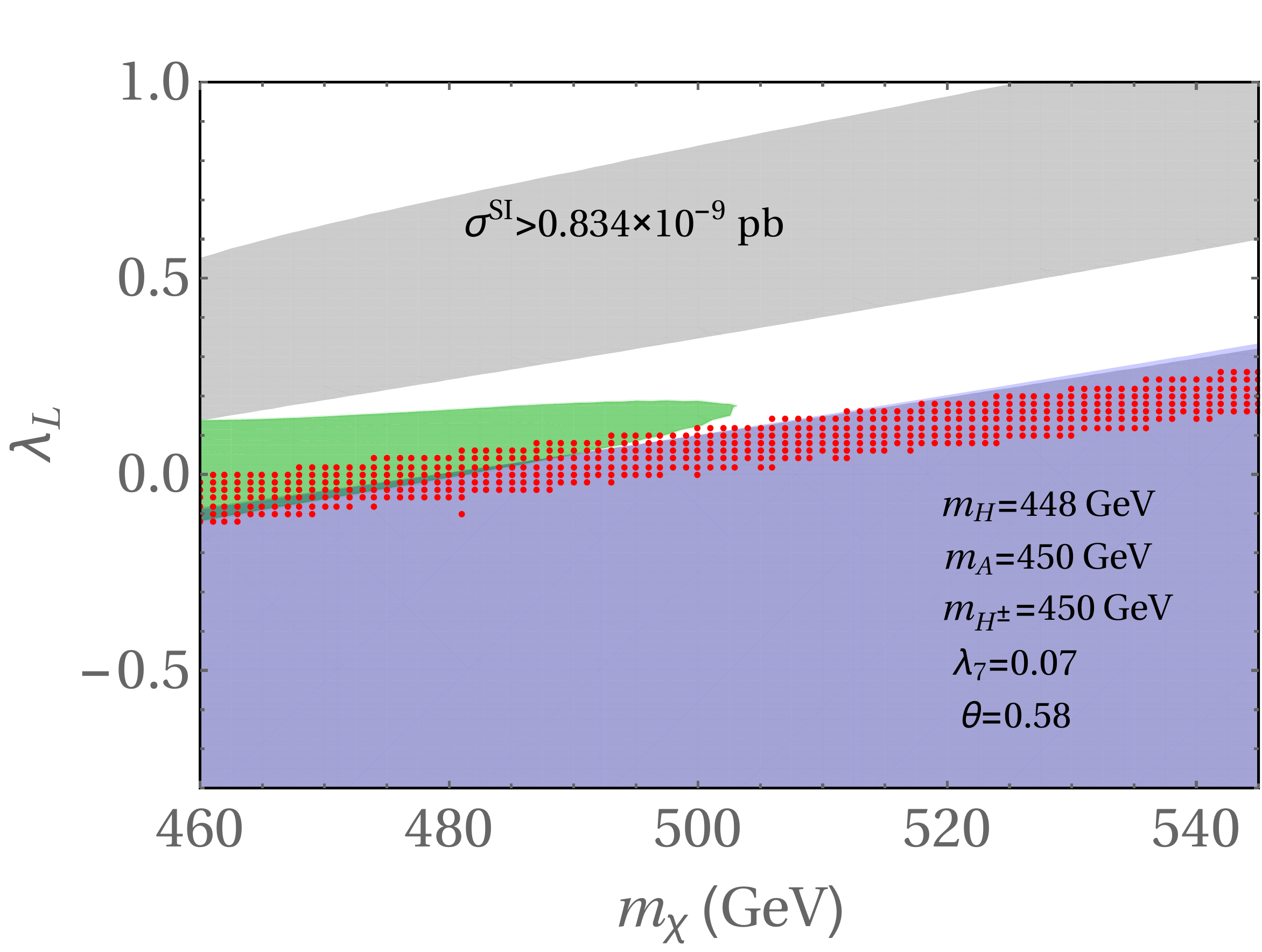}\\
\includegraphics[width=.4\textwidth]{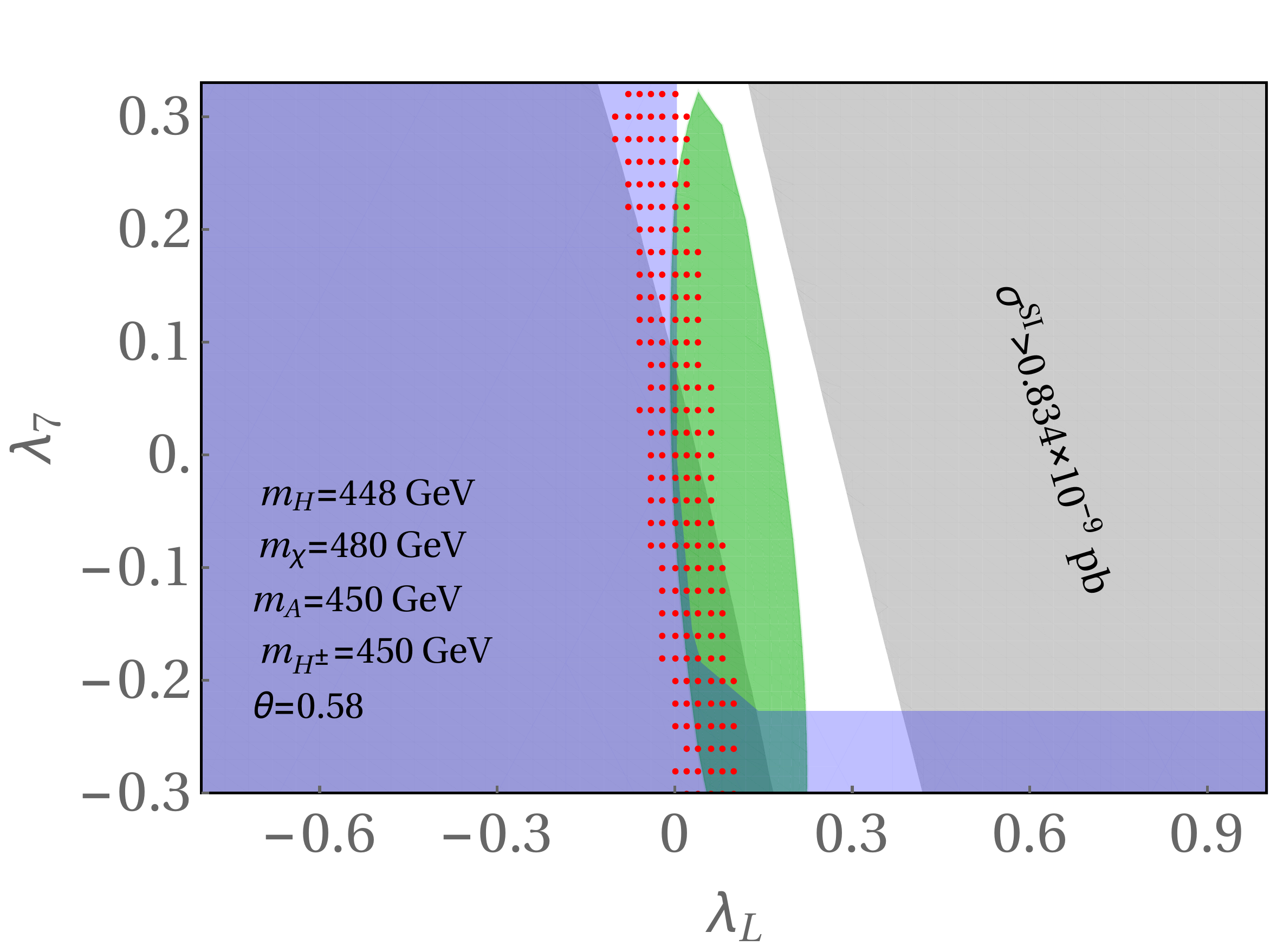}
\includegraphics[width=.4\textwidth]{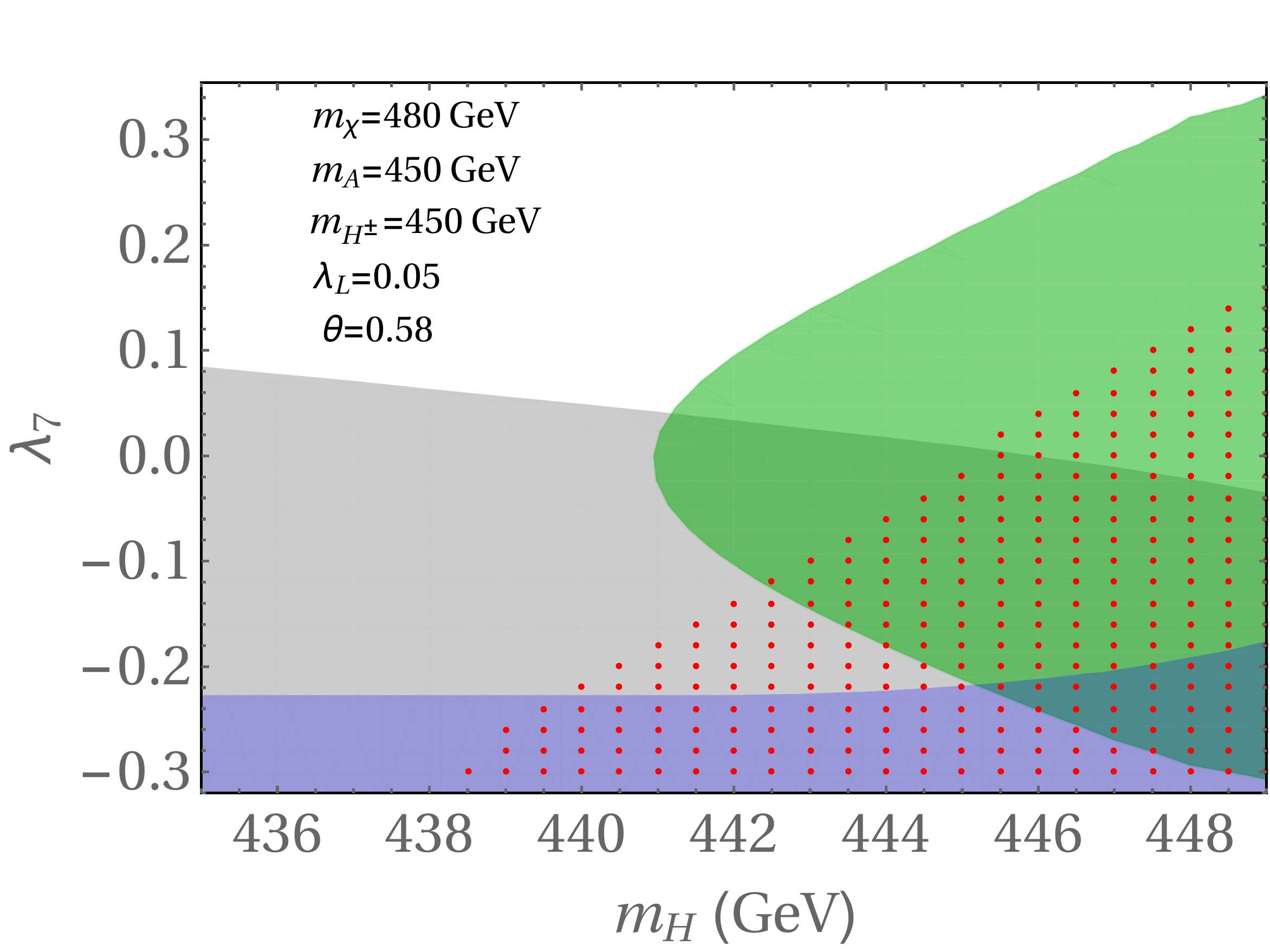}
\caption{For BP-c2 with fixed $\theta$, The red dotted regions stand for the parameter space can satisfy the condition $v_c/T_c>1$. The gray regions are excluded by LUX.  The blue shaded regions are excluded by the vacuum stability conditions.} 
\label{fig:results-BPcan2-2}
\end{centering}
\end{figure}

The combined results of Benchmark BP-c2 are presented in the Fig.~\ref{fig:results-BPcan2-1} and Fig.~\ref{fig:results-BPcan2-2}. In this benchmark model, the scalars $H$, $A$ and $H^{\pm}$ are nearly mass-degenerate.

\begin{enumerate}
\item {\bf EWPT}

To obtain the correct DM relic density the scalars in the inert doublet should be degenerate when the DM mass $m_H$ are larger than the gauge bosons', arising from the cancellation between different annihilation channels. However, the strength of the EWPT $v_c/T_c>1$ cannot be obtained due to the small mass spitting $\Delta M$, as found in Ref.~\cite{Blinov:2015vma}.
In this work, the existence of the extra singlet scalar $\chi$ provides an additional contribution to the vacuum barrier at finite temperature, as explored in Sec.~\ref{sec:PT}.
Therefore we get the chance to yield a strong enough EWPT as well as the correct DM relic abundance simultaneously\footnote{For the gravitational wave generated during the EWPT in the multi-scalar models we refer to Ref.~\cite{Bian:2017wfv,Chao:2017vrq}.}. 

We find that with the increase of the mass of the singlet scalar, the mixing angle needs to be smaller to meet the strong first order EWPT condition, as shown in 
 the top-left plot of Fig.~\ref{fig:results-BPcan2-1}. And also the criteria severely  constrain the coupling $\lambda_L$ as shown in the top-right plots of Fig.~\ref{fig:results-BPcan2-1} and bottom plots of Fig.~\ref{fig:results-BPcan2-2}.

\item{\bf DM relic density}

In the top-left panel of Fig.~\ref{fig:results-BPcan2-1} and top panels of Fig.~\ref{fig:results-BPcan2-2}, the mass difference of $m_\chi^2-m_H^2$ changes the DM-Higgs coupling $a_{hHH}$ efficiently (see Eq.~\ref{eq:scalarcouplings} and Eq.~\ref{eq:hHH}), leading to the variation of relic abundance. The correct DM relic density constrains the mass of the singlet scalar $\chi$ to be smaller than 500 GeV. 

From the top-right of Fig.~\ref{fig:results-BPcan2-1} and bottom panels of Fig.~\ref{fig:results-BPcan2-2}, we find that the relic density highly constrains on the parameters $\lambda_L$ to be small (around zero). The reason is that the coupling $a_{hHH}\sim 2v\lambda_L\cos^2\theta$ in this case, and annihilation cross section of DM pair can be suppressed by a smaller $\lambda_L$ in order to avoid the relic underabundance (as also discussed in Subsec.~\ref{H-DM}).

In the large mass spitting region of $m_\chi$ and $m_H$, the term $-\mu_{\text{soft}}\sin^22\theta$ becomes the dominant part of the coupling $a_{hHH}$. So, as the $a^2_{hHH}$ increasing, the annihilation cross section of DM becomes larger which causes the DM relic density deficiency, see the $m_\chi-\theta$, $m_\chi-\lambda_7$ and $m_H-\lambda_7$ plots in Fig.~\ref{fig:results-BPcan2-1} and Fig.~\ref{fig:results-BPcan2-2}. Also, the DM relic density behaviors in $m_\chi-\lambda_7$ and $m_H-\lambda_7$ planes are symmetrical with respect to $\lambda_7$, this is because the annihilation cross section $HH\rightarrow hh$ is proportional to $\lambda_7^2$.

\item{\bf DM direct detection}

At the relative large DM mass region, the direct detection is less constraining.
However, one still needs small $\lambda_L$ to suppress the Higgs portal coupling $a_{hHH}$, LUX results rule out the most space of parameter $\lambda_L$ in the $\lambda_L-\theta, m_\chi-\lambda_L, \lambda_L-\lambda_7$ spaces. 
The exclusion in the bottom-right panels of Fig.~\ref{fig:results-BPcan2-1} and Fig.~\ref{fig:results-BPcan2-2} with respect to the varying DM mass tells that plenty parameter space is needed to probe in the underground experiments of the next  generation.

\end{enumerate}

\section{Comments on leptonic and hadronic collider search}
\label{sec:cllider}
\subsection{Invisible Higgs decay}
For the case of $m_h>2m_{H,\chi}$, the invisible decay of the SM like Higgs needs to be taken into account:
\begin{eqnarray}
\label{eq:inv}
&\Gamma(h\rightarrow H\chi)=\frac{\sqrt{m_h^4-2 m_h^2 \left(m_H^2+m_\chi^2\right)+\left(m_H^2-m_\chi^2\right)^2} \left(2 \mu _{\text{soft}}  \cos2 \theta +2\left(\lambda _L- \lambda _7\right) v \sin 2\theta \right)^2}{64 \pi  m_h ^3} \; ,\\
&\Gamma(h\rightarrow HH)=\frac{\sqrt{m_h^4-4 m_h^2 m_H^2} \left(-2 \mu _{\text{soft}} \sin2 \theta+2\left(\lambda _L- \lambda _7\right) v \cos2 \theta +2\left(\lambda _L+ \lambda _7\right) v\right)^2}{128 \pi  m_h ^3}\; ,\\
&\Gamma(h\rightarrow\chi\chi)=\frac{\sqrt{m_h^4-4 m_h^2 m_\chi^2} \left(2 \mu _{\text{soft}} \sin2 \theta -2\left(\lambda _L- \lambda _7\right) v \cos2 \theta +2\left(\lambda _L+ \lambda _7\right) v\right)^2}{128 \pi   m_h ^3}\; .
\end{eqnarray}

ATLAS search~\cite{Aad:2015txa} sets bounds on the invisible Higgs decay branching ratio
Br$(h\to \text{inv})  < 28\% $ at the 95$\%$ CL, which has been considered in our choice of DM scenarios with $m_{\chi,H}\leq m_h/2$, i.e., in the benchmark models BP1, BP3, and BP6.

\subsection{Diphoton rate deviation}

As in IDM, the charged state $H^\pm$ leads to an additional contribution 
to the loop-induced $h\rightarrow \gamma\gamma$ and $\gamma Z$ rates.~\cite{Swiezewska:2012eh,Arhrib:2012ia}. 
 The $h\rightarrow\gamma\gamma$ 
rate can be obtained as~\cite{hhg,Branco:2011iw,Djouadi:2005gj,Spira:1995rr,Swiezewska:2012eh,Arhrib:2012ia}
\be
\Gamma (h\rightarrow \gamma\gamma)
= \frac{\alpha^2  G_F m_h^3}{128\sqrt{2}\pi^3}\left|
\mathcal{A}_{\rm SM} + \frac{\lambda_3 v^2}{2m^2_{H^\pm}} \mathcal{A}_0\left(\frac{m_h^2}{4m^2_{H^\pm}}\right)
\right|^2,
\ee
here, $\mathcal{A}_{\rm SM} \approx -6.56 + 0.08i$ is the leading contribution coming
from $W$ bosons and top quarks for $m_h=125 $ GeV. 
The contribution coming from $H^\pm$ can enhance or suppress the $h\rightarrow \gamma\gamma$ rate relative to the 
SM depending on the sign of $\lambda_3$, which is characterized by the mass splitting of $m_H-m_{H^\pm}$ as well as the mixing of the singlet and inert doublet scalars. For our benchmarks where strong first order EWPT and correct DM relic abundance can be accomplished, an tinny enhancement of $h\rightarrow \gamma\gamma$ (around $\mathcal{O}(10^{-3})$) is obtained due to $\lambda_3 < 0$. Other benchmark scenarios favor $\lambda_3 > 0$, which may lead to a large suppression of the rate, around $\mathcal{O}(10^{-2})$. And the rate can be diluted by the invisible decay of SM-like Higgs when $m_{H,\chi}<m_h$,
as also noticed in Refs.~\cite{Borah:2012pu,Cline:2013bln, Blinov:2015vma}.

\subsection{Triple Higgs couplings v.s. EWPT}

\begin{figure}[!htbp]
\begin{centering}
\includegraphics[width=.4\linewidth]{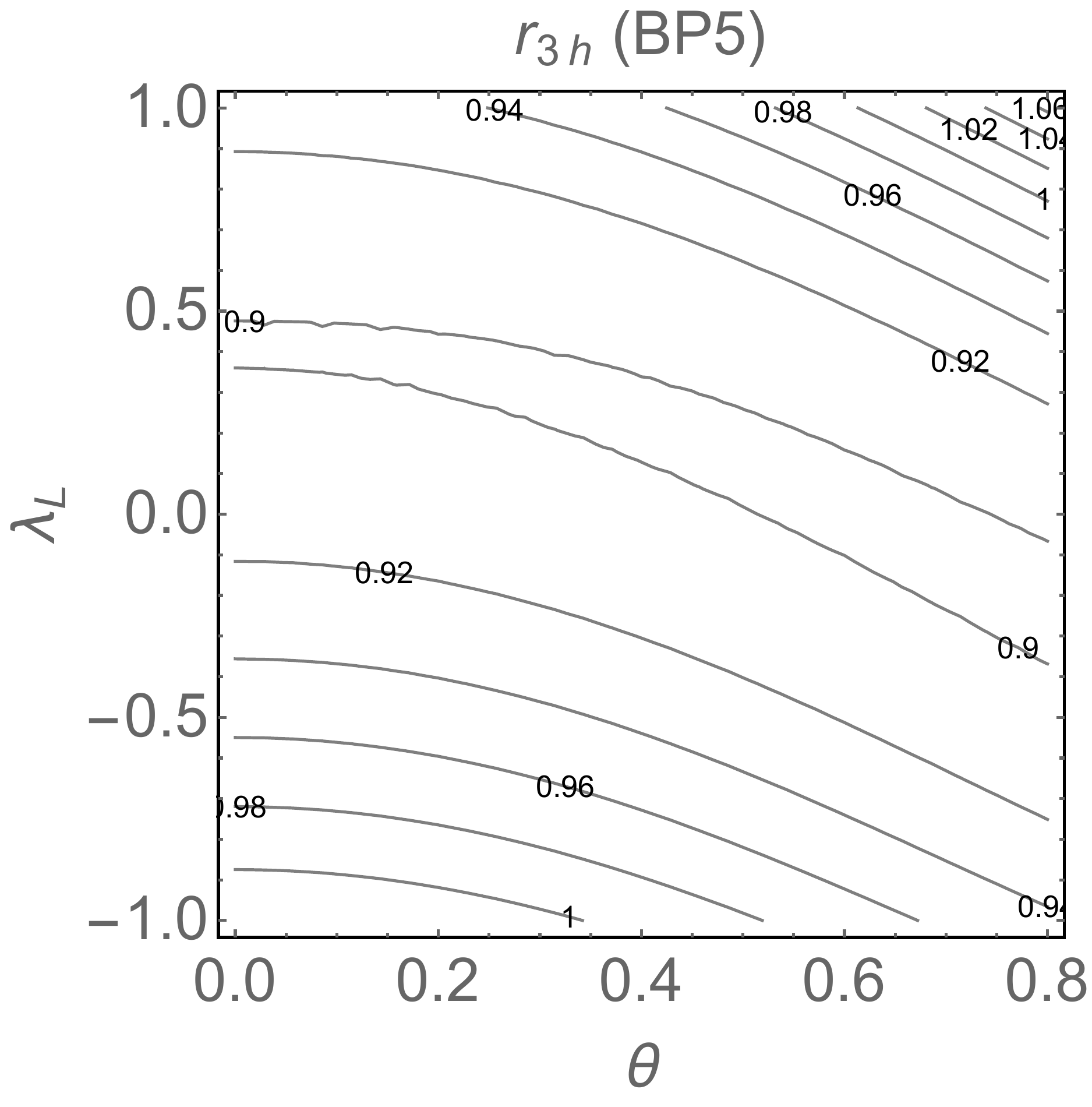}
\includegraphics[width=.4\linewidth]{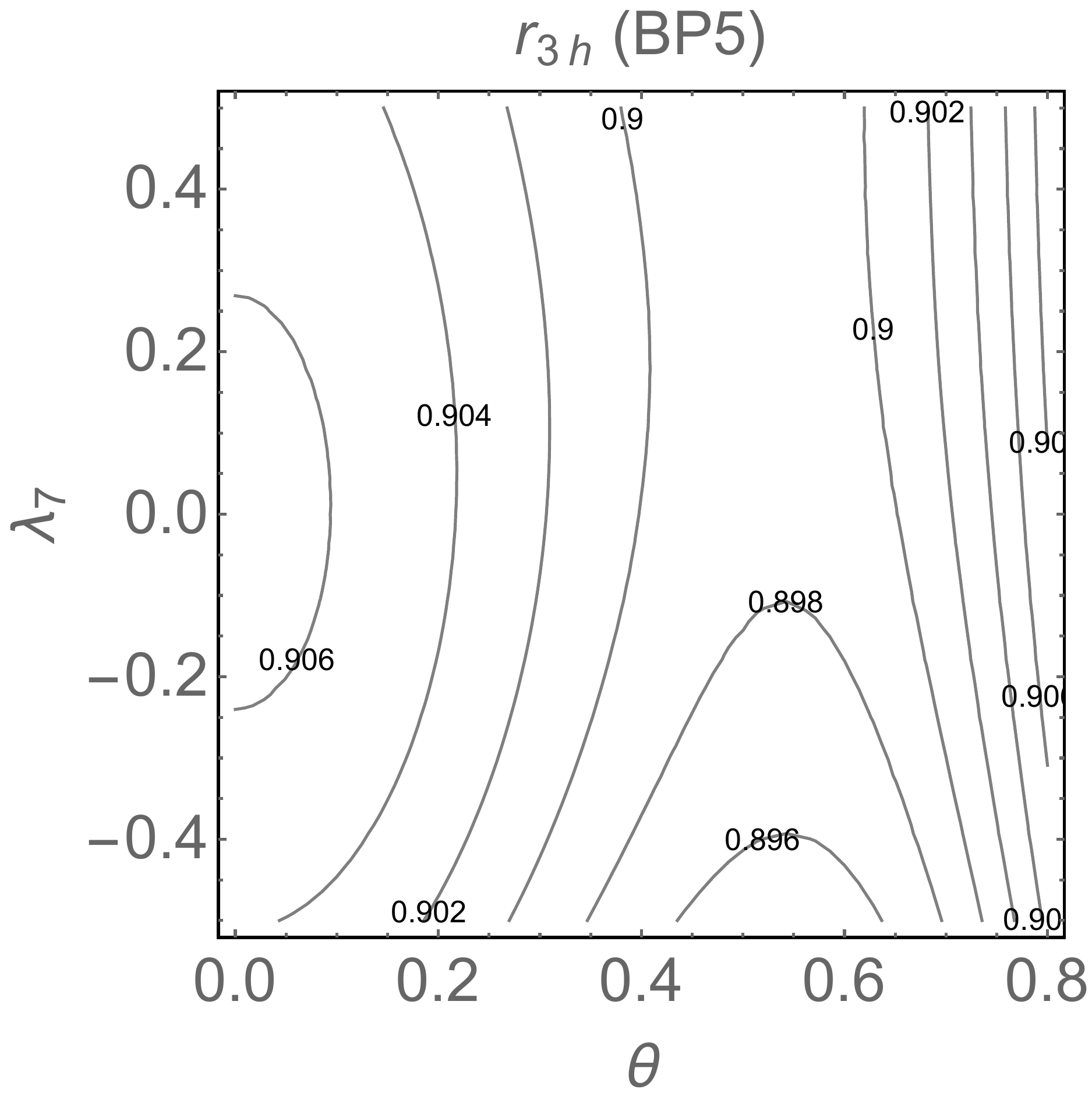}\\
\includegraphics[width=.4\linewidth]{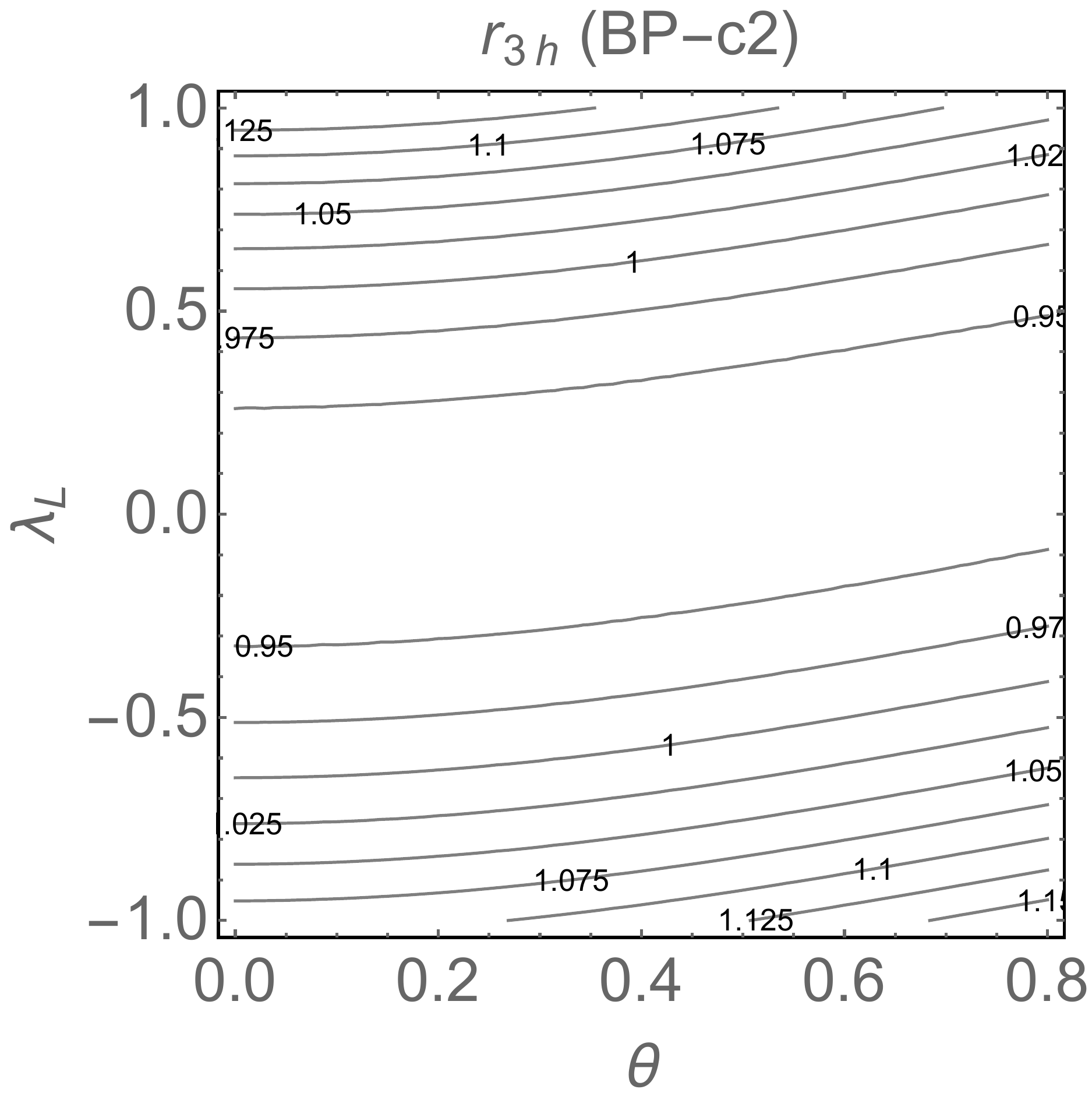}
\includegraphics[width=.4\linewidth]{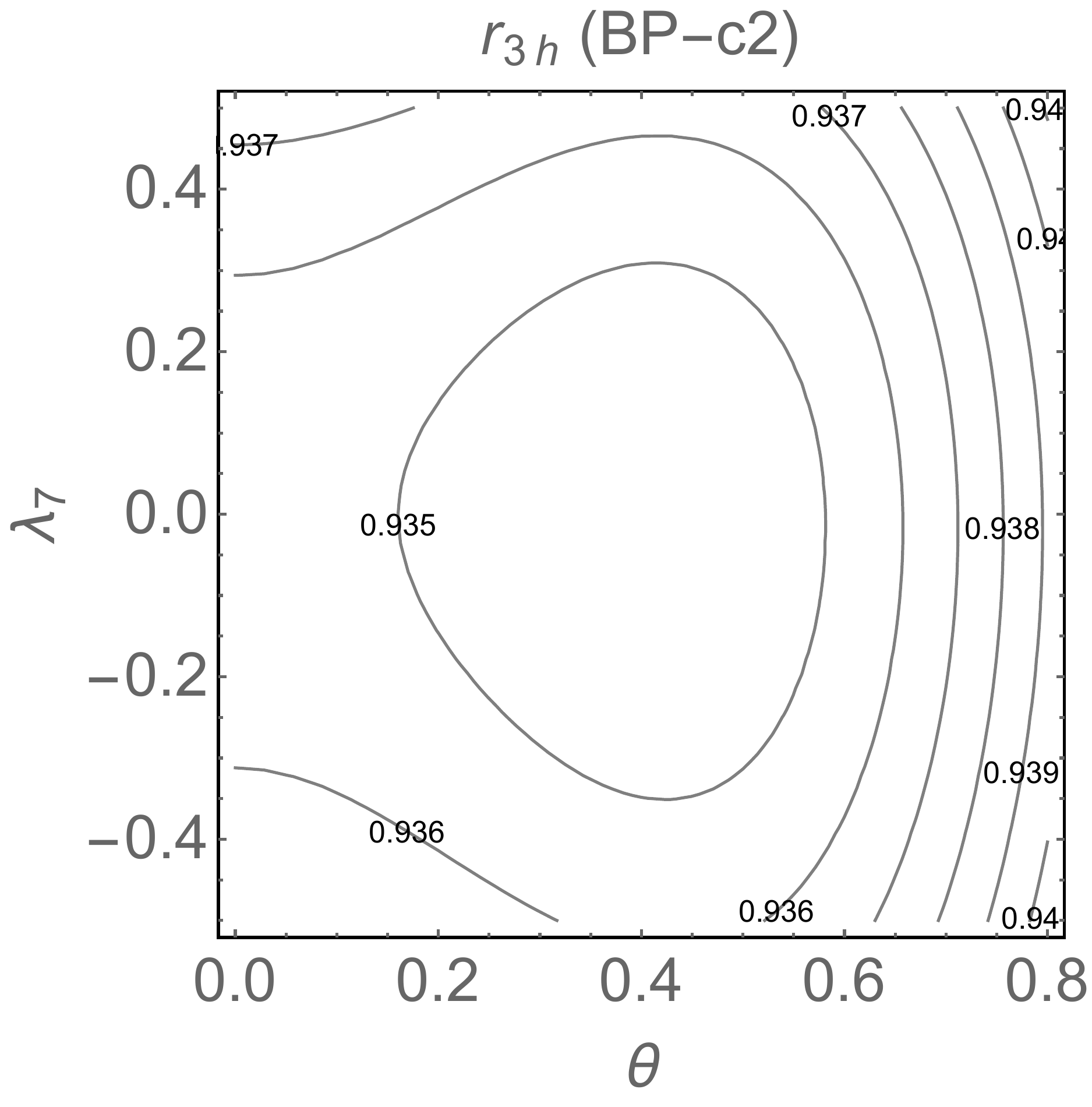}
\caption{Triple Higgs coupling ratio for BP5 and BP-c2 with respect to the mixing parameter $\theta$, $\lambda_L$ and $\lambda_7$.} 
\label{fig:3h1}
\end{centering}
\end{figure}

As mentioned in the Section~\ref{section:introduction}, the EWBG stands out from many baryogenesis mechanisms. 
Based on the mechanism, the triple Higgs couplings might be modified which can be probed at the future hadron colliders, such as SPPC~\cite{Arkani-Hamed:2015vfh}.
However, the strong first order phase transition not necessarily requires the enhancement or suppression of the triple Higgs coupling comparing with SM prediction, as studied in the singlet assisted case in Ref.~\cite{Profumo:2014opa}. 
Since there is no mixing between the SM-like Higgs and other neutral Higgses, we expect the model does not 
cause large modification of the triple Higgs couplings in 
the allowed benchmark scenarios. Generally, we can expect a tinny loop correction to the triple Higgs couplings ($C_{hhh}$) from new scalar sectors.
The ratio of $C_{hhh}$ of our model with respect to the SM can be obtained as
\begin{equation}
r_{3h}=\frac{C_{hhh}}{C^{SM}_{hhh}}\;,
\end{equation}
with $C_{hhh}=d^3{V_{eff}}(T=0)/d{h^3}$. The plots of ratio $r_{3h}$ in Fig.~\ref{fig:3h1} indicates that the deviation from the SM prediction is around $\mathcal {O}(10^{-2}-10^{-1})$ in the benchmark models  BP5 and BP-c2. When the triple Higgs vertices being modified by the magnitude of $\mathcal {O}(10^{-1})$, which falls in the projection of the ILC~\cite{Asner:2013psa}\footnote{For the EWPO search of the model at CEPC, we refer to Ref.~\cite{Cai:2017wdu}.}, one can expect the benchmark modes being detected through the process of $e^+e^-\rightarrow  Zhh$. When the deviation is $\sim \mathcal {O}(10^{-2})$, one may expect the benchmark models to be probed by
the SPPC~\cite{Arkani-Hamed:2015vfh,Yao:2013ika} through the process of $pp\rightarrow hh$. 

\subsection{Lepton collider and hadron collider prospects }
\label{sec:LPP}
\begin{figure}[!htbp]
\begin{centering}
\includegraphics[width=.4\linewidth]{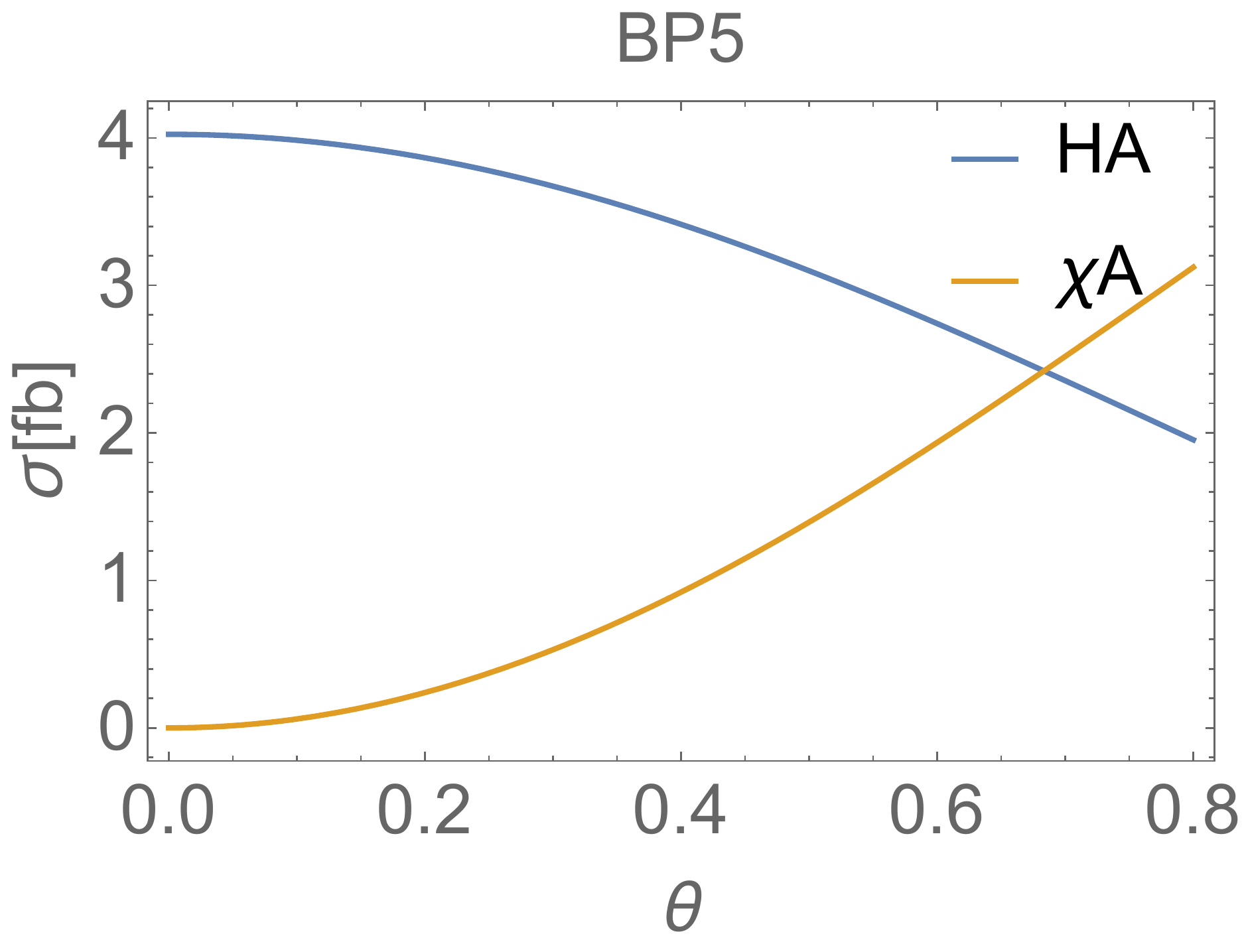}
\includegraphics[width=.4\linewidth]{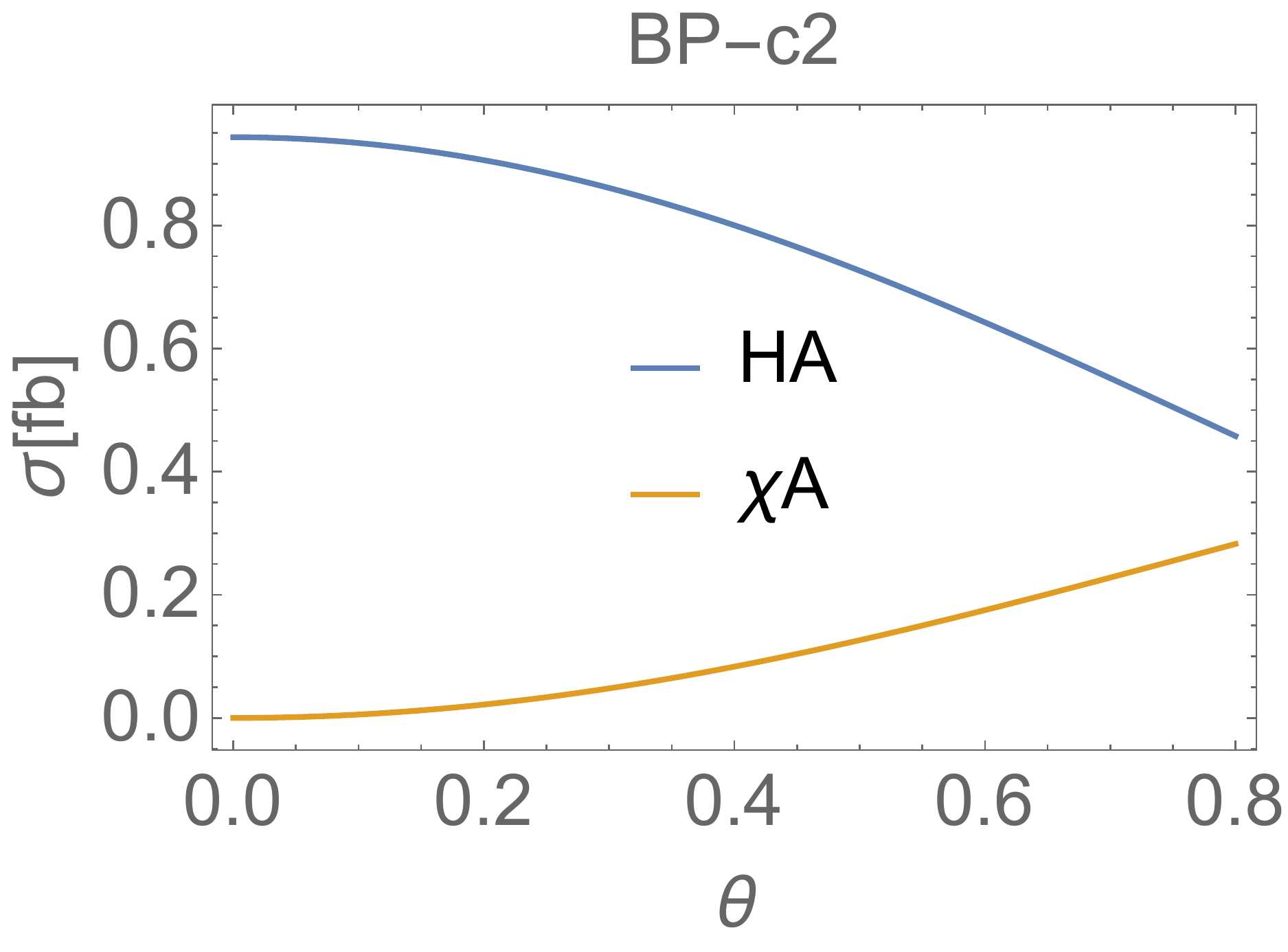}
\caption{The cross section of $\sigma(e^+e^-\rightarrow H(\chi)A)$ for BP5 and BP-c2 with respect to the mixing parameter $\theta$.} 
\label{fig:ilc}
\end{centering}
\end{figure}

At first, due to the mixing of $\chi$ and $H$ in the model, one can expect the 
pair production process of $H(\chi)A$ at leptonic colliders~\cite{Barbieri:2006dq,Drees:1989uu}, the corresponding cross section is 
\begin{align}\label{eq:lc}
\sigma(e^+e^-\rightarrow H(\chi)A)&=g_{H(\chi)AZ}^2
\frac{2\pi\alpha_{em}^2((-\frac{1}{4})^2+(-\frac{1}{4}+\sin^2\theta_w)^2)}{3\sin^4\theta_w\cos^4\theta_w}\nonumber\\
 &\times\frac{\lambda^3_{H(\chi)A}}{\sqrt{s}\left((s-m_Z^2)^2+m_Z^2\Gamma_Z^2\right)}\;,
\end{align}
where $g_{H(\chi)AZ}=\cos\theta(\sin\theta)$ and $\lambda_{ij}=\left[\left(s-M_i^2-M_j^2)^2-4 M_i^2M_j^2\right)/4s\right]^{1/2}$ is the space function of the two-body phase space. 
 Unfortunately, for our two benchmark models (BP5 and BP-c2) that can address both the strongly first order EWPT and DM relic abundance, the future leptonic colliders, such as CEPC~\cite{cepc}, and FCC-ee~\cite{Gomez-Ceballos:2013zzn}, lack the ability to search the $H(\chi)A$ pair productions, since such processes are kinematically unaccessible for the projected design of the centre-of-mass energy. We plot the cross sections predictions at $\sqrt{s}=1000$ GeV at ILC for the two benchmark models in Fig.~\ref{fig:ilc}. With the increasing of the mixing angle $\theta$, one obtain the increasing (decreasing) of the cross section of $e^+e^-\rightarrow \chi(H)A$. 

As in the IDM, the added scalars may alleviate the Higgs natrualness problem since more bosonic fields contributions are added to the
Veltman Condition~\cite{Barbieri:2006dq,Bian:2014cja}. While for the naturalness problem to be probed at leptonic colliders, one needs to expect a sizable modification of wave-function renormalization of the SM Higgs field after these extra Higgses and dark matter fields ($H,\chi,A$) are integrated out~\cite{Craig:2013xia,Curtin:2014jma,Huang:2017rzf}.  We leave the detailed study on interplay between naturalness probe and EWPT to the future study.

At the Hadronic Collider, the dominant production channels of the extra Higgs pairs in the model can be\footnote{The production channel involving an off-shell $h$ is negligible here.}
\begin{align}
pp  &  \rightarrow W^{\ast}\rightarrow H^{\pm}A\text{ or }H^{\pm}H(\chi)\label{HS}\; ,\\ 
pp  &  \rightarrow Z^{\ast}(\gamma^{\ast})\rightarrow H(\chi)A\text{ or }H^{+}H^{-}\; ,\label{wzdec}%
\end{align}
and followed by the decay channels being%
\begin{align}
H&\rightarrow h\chi\;, \\
H^\pm  &  \rightarrow  \chi W^\pm\label{Hd}\; ,\\
A  &  \rightarrow \chi Z^{\ast}\; . \label{Ad}%
\end{align}
We explore the possible prospects of the future search of the two benchmark models: BP5 and BP-c2. We use {\tt MadGraph5}~\cite{Alwall:2014hca} to estimate cross sections of different channels, the cross section of the $H^\pm A$, $H^\pm H$ and $H^\pm\chi$ channels of Eq.~\ref{HS} are estimated to be 2.9 fb, 2.1 fb and 0.6 fb at 14 TeV, the cross section of the $HA$, $\chi A$ and $H^{+}H^{-}$ channels of Eq.~\ref{wzdec} are estimated to be 1.6 fb, 0.5 fb and 0.7 fb. As in the usual IDM models, the multi-lepton plus missing energy signatures~\cite{Miao:2010rg,Gustafsson:2012aj,Belanger:2015kga} can be expected in the final states at the detector level.
While, due to the smallness of the cross sections, these channels might be unreachable by LHC run-2 at 14 TeV or even SPPC.
More detailed studies are left to the future projects.

As for the BP-c2, the scalar mass relation of $m_{H^\pm}=m_A>m_H$ indicate that the followed decay of Eq.~\ref{Hd} is replaced by $H^\pm\rightarrow HW^{\ast}$, the macroscopic decay lengths of which can be $c\tau\sim 1 mm$~\cite{Blinov:2015sna}  for the mass splitting between $H^\pm(A)$ and $H$ being 2 GeV.
And the bottom-right panels of the Fig.~\ref{fig:results-BPcan2-1} and Fig.~\ref{fig:results-BPcan2-2} illustrate that $m_H$ can be very close to and even degenerate with the $m_{H^\pm,A}$  for $\theta>0.5$ and $\lambda_7>0.07$, which means one can expect $c\tau> 1 mm$ and therefore the displaced decay of $H^\pm\rightarrow HW^{\ast}$ and 
$A\rightarrow H Z^{\ast}$  can be searched  through a mono-jet with a soft displaced vertex~\cite{Bai:2011jg,Weiner:2012cb,Izaguirre:2015zva}. This signature can make the benchmark model different from the IDM case~\cite{Blinov:2015sna}. When the $\lambda_L\leq -0.2$, one have $m_H\sim m_{A,H^\pm}$ for dark matter relic abundance undersaturated case of BP5. Then, one can also expect the displaced vertex signature.

\section{Conclusion and discussion}
\label{sec:con}

In this work, we study the dark matter and electroweak phase transition in the framework of mixed scalar dark matter model. In the model,
the dark matter candidate is the  lightest $\Z_2$-odd mixed particle coming from  the mixing of the singlet scalar and the CP-even neutral component of the inert doublet.
We investigate the scenarios where the dark matter particle is mostly the CP-even neutral component of the inert doublet (the benchmark scenario with $H$ being the DM candidate) or the singlet (with $\chi$ being the DM candidate), which is determined by the mixing of the two. After imposing theoretical constraints of unitarity, stability and electroweak precision test(in particular the $T$ parameter), we explore several benchmark models classified by the mass of the dark matter particle. 
The dark matter annihilation process can be affected a lot by the mixing and co-annihilation effects.
Two benchmark models in which the strongly first order electroweak phase transition can be accomplished have been investigated. 

The effects of quartic scalar couplings ($\lambda_L, \lambda_7$), DM mass, mass splittings among different extra scalars, and mixing angle on the phase transition as well as dark matter phenomenology have been studied. 
Certain mass differences between the mixed scalars $\chi, H$ and other scalars in the inert doublet ($H^\pm$ and $A$) are required to successfully realize the strongly first order electroweak phase transition. 
The DM direct detection limits can be evaded through tuning the mixing parameter. 
Lepton and hadron collider prospects have also been addressed in the model.

The unbroken $\Z_2$ symmetry in the model precludes spontaneous and explicit CP violation arising from the Higgs potential, because of which the model fails to accommodate successful electroweak baryogenesis to explain the baryon asymmetry of the universe.  One remedy method could be the introduction of effective high dimensional operators~\cite{Dine:1990fj,Dine:1991ck}, which is beyond the scope of this study.

\section*{Acknowledgments}
\vspace*{-2mm}

We would like to thank the Institute of Theoretical Physics at the Chinese Academy of Sciences for their hospitalities when part of this work was prepared.  The work of LGB is partially supported by the National Natural Science Foundation of China (under Grant No. 11605016), Basic Science Research Program through the National Research Foundation of Korea (NRF) funded by the Ministry of Education, Science and Technology (NRF-2016R1A2B4008759), and Korea Research Fellowship Program through the National Research Foundation of Korea (NRF) funded by the Ministry of Science and ICT (2017H1D3A1A01014046). The work of X.Liu is supported by the China Postdoctoral Science Foundation (under Grant No.\ BX201700116).


\vspace*{8mm}
\appendix
\section{Thermal field masses}
\label{app:sec}

The transverse components of $W$ boson do not receive thermal corrections, whereas the longitudinally polarized $W$ boson are corrected  as
\be
m_{W_L}^2 = m_W^2 + 2 g^2 T^2.
\ee 

The masses of the longitudinal $Z$ and $A$ are determined by diagonalizing the matrix 
\be
\frac{1}{4}h^2
\begin{pmatrix}
g^2& -g g^\prime \\
-g g^\prime & g^{\prime 2}
\end{pmatrix}
+
\begin{pmatrix}
2 g^2 T^2 & 0 \\
0 & 2 g^{\prime 2} T^2 
\end{pmatrix}.
\ee
The eigenvalues can be written as
\be
m_{Z_L,A_L}^2 = \frac{1}{2} m_Z^2 + (g^2 + g^{\prime 2} )T^2 \pm \Delta, 
\ee
where 
\be
\Delta^2 = \left(\frac{1}{2} m_Z^2 + (g^2 + g^{\prime 2} )T^2 \right)^2
- g^2 g^{\prime 2} T^2 ( h^2 + 4 T^2). 
\ee
Following the approach of Ref.~\cite{Carrington:1991hz}, the field dependent thermal masses of scalar masses are calculated to be:
\begin{eqnarray}
m_{hh}(T)&=&(m_h^2-3\lambda_1 v^2)+\frac{T^2}{12}(6\lambda_1+2\lambda_3+\lambda_4+(9 g^2+3g'^2)/4+3g_t^2+\lambda_7)+3[m_h^2/(2v^2)] h^2\, ,\nonumber\\
m_{GG}(T)&=&(m_h^2-3\lambda_1 v^2)+\frac{T^2}{12}(6\lambda_1+2\lambda_3+\lambda_4+(9 g^2+3g'^2)/4+3g_t^2+\lambda_7)+3[m_h^2/(2v^2)] h^2\, ,\nonumber\\
m_{SS}(T)&=&2\mu_S^2+\lambda_7 h^2+\frac{T^2}{24}(12 \lambda_6+4\lambda_7+4\lambda_8)\, ,\nonumber\\
m_{H_0}(T)&=&\mu_I^2+2\lambda_L h^2+\frac{T^2}{12}(6\lambda_2+2\lambda_3+\lambda_4+\lambda_8)\, ,\nonumber\\
m_{A_0}(T)&=&\mu_I^2+\frac{T^2}{12}(6\lambda_2+2\lambda_3+\lambda_4+(9 g^2+3g'^2)/4+\lambda_8)+\fr{1}{2}\lambda_A h^2\, ,\nonumber\\
m_{H^{\pm}}(T)&=&\mu_I^2+\frac{T^2}{12}(6\lambda_2+2\lambda_3+\lambda_4+(9 g^2+3g'^2)/4+\lambda_8)+\fr{1}{2}\lambda_3 h^2\, ,
\end{eqnarray}
with $\mu_{I,S}^2$ given by Eq.~\ref{eq:scalarcouplings}, and 
$g_t=\sqrt{2}m_t/v$.


\bibliographystyle{JHEP}

\end{document}